\documentclass[aps,twocolumn,superscriptaddress,amsmath]{revtex4-1}

\pdfoutput = 1

\usepackage[pdftex]{graphicx,color}
\usepackage{soul}
\usepackage{amsfonts}
\usepackage{amssymb}
\usepackage{amsmath}
\usepackage{amsthm}
\usepackage{bm}
\usepackage{braket}
\usepackage{ulem}
\usepackage{enumerate}

\usepackage{subfigure}
\usepackage{rotate}
\usepackage{bm}
\usepackage{dcolumn}

\usepackage[pdftex]{hyperref}

\let\oldmarginpar\marginpar
\renewcommand\marginpar[1]{\-\oldmarginpar[\raggedleft\tiny #1]
{\raggedright\tiny #1}}

\newcommand{\argmax}{\operatornamewithlimits{arg\ max}}

\graphicspath{{figs/}}

\begin{document}

\title{Quantum algorithm for energy matching in hard optimization problems}

\author{C. L. Baldwin}
\affiliation{Department of Physics, Boston University, Boston, MA 02215, USA}
\affiliation{Department of Physics, University of Washington, Seattle, WA 98195, USA}

\author{C. R. Laumann}
\affiliation{Department of Physics, Boston University, Boston, MA 02215, USA}

\date{\today}

\begin{abstract}
We consider the ability of local quantum dynamics to solve the ``energy matching'' problem:
given an instance of a classical optimization problem and a low energy state, find another macroscopically distinct low energy state.
Energy matching is difficult in rugged optimization landscapes, as the given state provides little information about the distant topography.
Here we show that the introduction of quantum dynamics can provide a speed-up over local classical algorithms in a large class of hard optimization problems. 
The essential intuition is that tunneling allows the system to explore the optimization landscape while approximately conserving the classical energy, even in the presence of large barriers.
In particular, we study energy matching in the random $p$-spin model of spin glass theory.
Using perturbation theory and numerical exact diagonalization, we show that introducing a transverse field leads to three sharp dynamical phases, only one of which solves the matching problem:
(1) a small-field ``trapped'' phase, in which tunneling is too weak for the system to escape the vicinity of the initial state;
(2) a large-field ``excited'' phase, in which the field excites the system into high energy states, effectively forgetting the initial low energy;
and (3) the intermediate ``tunneling'' phase, in which the system succeeds at energy matching.
We find that in the tunneling phase, the time required to find distant states scales exponentially with system size but is nevertheless exponentially faster than simple classical Monte Carlo.
\end{abstract}

\maketitle

\tableofcontents

\section{Introduction} \label{sec:introduction}

\begin{figure}[t]
\begin{center}
\includegraphics[width=0.9\columnwidth]{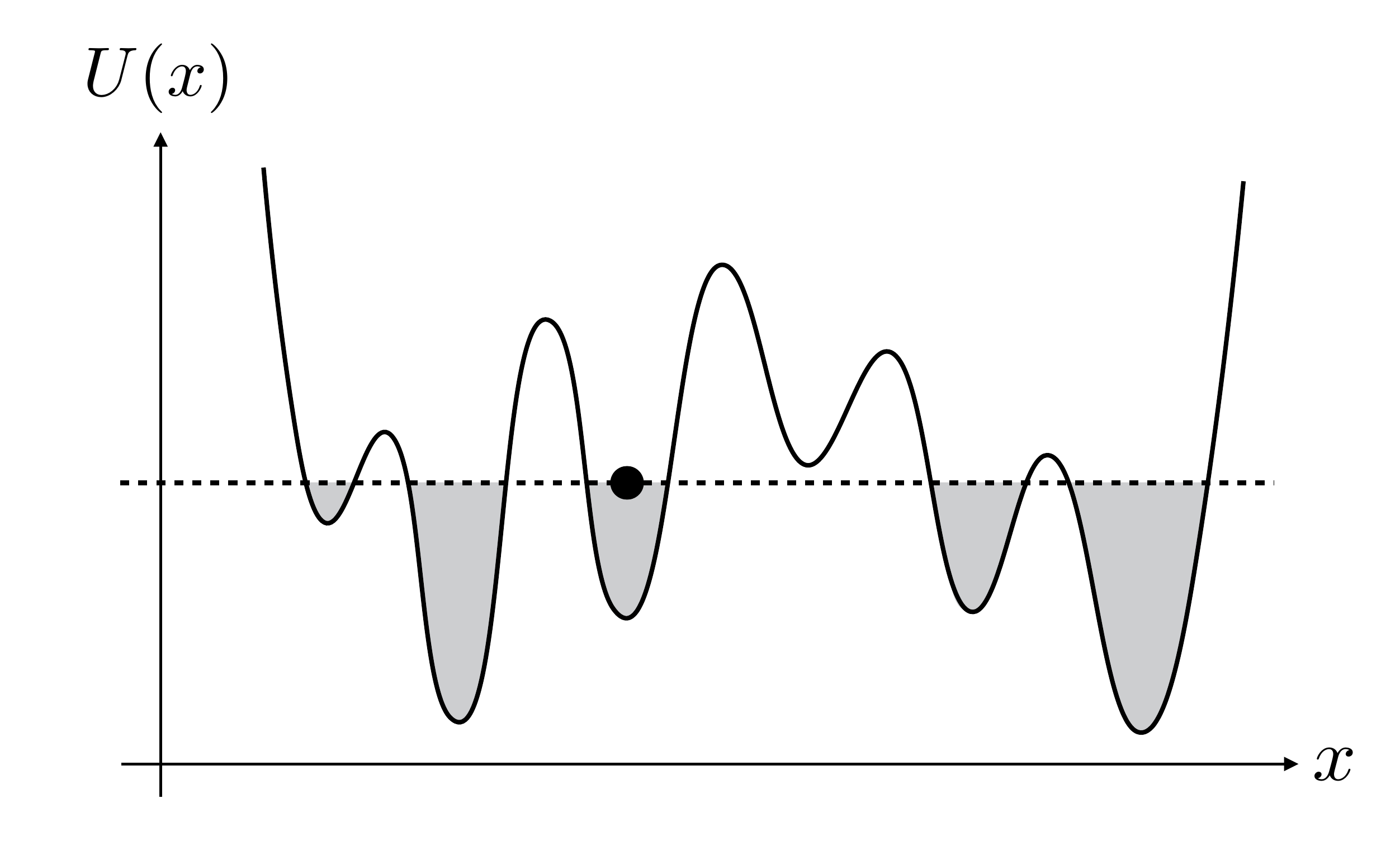}
\includegraphics[width=1.0\columnwidth]{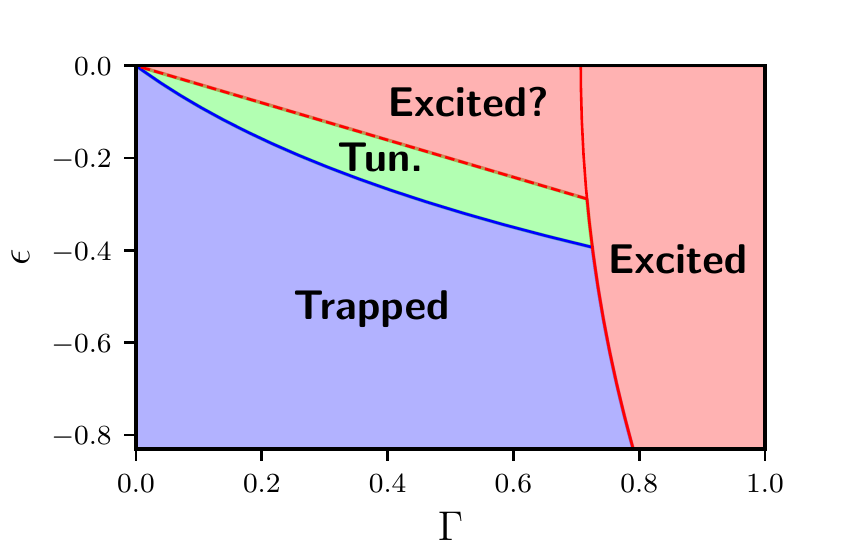}
\caption{(Top) A one-dimensional example of a rugged energy landscape $U(x)$. States with energy below the dashed line form disconnected clusters (shaded). A particle (black dot) remains in a cluster until it either surmounts or tunnels through the energy barriers. (Bottom) The dynamical phase diagram of the Random Energy Model ($p \rightarrow \infty$ limit). The system tunnels between clusters in the ``tunneling'' phase (green), remains trapped in a cluster in the ``trapped'' phase, and is excited out of clusters in the ``excited'' phase. The trapped, tunneling, and large-$\Gamma$ excited phases have been confirmed numerically. It is unclear whether the small-$\Gamma$ excited phase is truly an excited phase or a portion of the tunneling phase in which perturbation theory does not apply.}
\label{fig:REM_phase_diagram}
\end{center}
\end{figure}

Suppose that the perennial traveling saleswoman, having many responsibilities herself, delegates the menial task of constructing an efficient sales route to the company intern.
The intern wants to impress his boss, and decides to produce not one but many routes, each sufficiently distinct from the others so that the saleswoman may pick the one which suits her best.
After laboring for days on the notoriously hard traveling saleswoman's problem~\cite{Cook2012Pursuit}, the intern succeeds at last in identifying one efficient route, but he shudders at the thought of having to repeat the process multiple times.
Can he make use of this first route to construct others faster than he would from scratch, keeping in mind that the additional routes must be sufficiently different?
This is the matching problem: given an optimization problem and one optimal or near-optimal solution, find others that are sufficiently distinct.
The starting solution serves as a hint for finding the others.
Matching is relevant both when the original optimization problem is difficult~\cite{Mezard2002Analytic,Mulet2002Coloring,Raymond2007Phase} and when efficient algorithms are only capable of finding a small set of special solutions~\cite{Krzakala2009Hiding,Baldassi2016Unreasonable}.

The matching problem is often difficult for the same reason as the original problem: ruggedness in the cost function to be minimized.
A one-dimensional example of a rugged cost function (or equivalently, potential energy) is sketched in the top panel of Fig.~\ref{fig:REM_phase_diagram}, and provides intuition for the high-dimensional configuration spaces of real optimization problems.
If the energy landscape has deep local minima, called ``clusters'', then a starting configuration in one cluster does not provide any benefit for finding distant clusters of solutions.
Local search algorithms such as Metropolis Monte Carlo must excite the system out of clusters in order to explore the configuration space.
Furthermore, it is straightforward to show that certain matching problems are NP-complete, and we provide a short proof in Appendix~\ref{appendix:np_complete}.

In the present paper, we assess whether \textit{quantum Hamiltonian} dynamics may be faster than classical algorithms at energy matching in rugged landscapes.
We take as a non-trivial testbed the classical random $p$-spin model of spin glass theory. 
This model has sharply defined clusters of low energy states (as reviewed in more detail in Sec.~\ref{sec:p_spin_model}). 
We denote the $p$-spin Hamiltonian by $H_p$, which is diagonal in the $\hat{\sigma}^z$ basis of $N$ spin-1/2s.
Quantum dynamics is produced by applying a uniform transverse field:
\begin{equation} \label{eq:Hamiltonian_form}
H = H_p - \Gamma \sum_{i=1}^N \hat{\sigma}_i^x.
\end{equation}
Starting in a classical (i.e., $\hat{\sigma}^z$) state $\ket{\sigma}$, we study the probability for observing at time $t$ a classical state $\ket{\sigma'}$ belonging to a different cluster, i.e., $\big| \braket{\sigma' | e^{-i H t} | \sigma} \big| ^2$.
Hamiltonians of the form in Eq.~\eqref{eq:Hamiltonian_form} have long been used in the context of quantum computation, particularly with time-dependent coefficients to study the quantum adiabatic algorithm~\cite{Farhi2001Quantum,Knysh2008Statistical,Young2010First,Farhi2012Performance,Isakov2016Understanding} for finding ground states.
Here, we instead use a static Hamiltonian to study energy matching dynamics, and do not restrict ourselves to ground states.
Our analysis applies to approximate as well as perfect optimization.

One might expect quantum dynamics to be efficient for two reasons: conservation of energy biases the dynamics towards classical states having the same energy as the starting configuration, and quantum fluctuations can tunnel through the energy barriers that separate those states.
However, Hamiltonian dynamics conserves the full quantum mechanical energy $\braket{H}$, whereas the goal of energy matching is to find a state with the same classical energy $\braket{H_p}$.
Furthermore, recent work has shown that the tunneling amplitudes between clusters can be exponentially suppressed in many-body systems~\cite{Altshuler2010Anderson,Baldwin2016Many,Baldwin2017Clustering}.
Thus the performance of quantum dynamics in energy matching, and its comparison to classical search algorithms, is non-trivial.

We find three sharp dynamical phases for the transverse field $p$-spin models, each with distinct implications for energy matching.
The phase depends on the target energy per spin $\epsilon$ and the strength of the transverse field $\Gamma$.
A representative phase diagram is shown in the bottom panel of Fig.~\ref{fig:REM_phase_diagram}.
At low $\epsilon$ and small $\Gamma$, the probability of the system tunneling between clusters vanishes in the thermodynamic limit even at arbitrarily late times.
The system cannot exit its initial cluster and energy matching fails in this ``trapped'' phase.
At large $\Gamma$, the system moves freely out of the initial cluster but is excited to higher classical energies in return for magnetizing along the transverse field.
Energy matching fails in this ``excited'' phase as well, since the system does not locate states at the desired classical energy.
Only at intermediate $\epsilon$ and $\Gamma$, in the ``tunneling'' phase, does energy matching succeed by tunneling between clusters while roughly preserving the classical energy density~\footnote{Dynamical fluctuations in the classical energy density are $O(1/N)$ in the REM, and are $O(1/p)$ in the $p$-spin model.}.
The timescale for tunneling is exponential in system size, i.e., quantum dynamics cannot solve the matching problem in polynomial time.
On the other hand, classical algorithms also require exponential runtime in these models; quantum dynamics runs exponentially \textit{faster} than simple classical Monte Carlo simulations.

Note that the tunneling phase does not exist at sufficiently low $\epsilon$ for the $p$-spin model.
A uniform transverse field cannot solve the matching problem near the classical ground state regardless of the field strength and regardless of runtime.

We derive these results using both perturbation theory in $\Gamma$ and numerical exact diagonalization.
Here we present the underlying intuition.
We work in the $\hat{\sigma}^z$ basis, whose eigenstates are referred to as classical states and have definite classical energy $H_p$.
Since matching problems start from a given classical state $\ket{\sigma}$ with specified energy density $\epsilon$, we distinguish states $\ket{\sigma'}$ not only by their classical energy densities $\epsilon'$ but also their fractional Hamming distances $x$ relative to $\ket{\sigma}$:
\begin{equation} \label{eq:fractional_distance_def}
x \equiv \frac{1}{2} \left( 1 - \frac{1}{N} \sum_{i=1}^N \sigma_i \sigma'_i \right) .
\end{equation}
States with $\epsilon' = \epsilon$, assuming $\epsilon$ is sufficiently low, are disconnected: some states lie at distances less than a certain $x^*(\epsilon)$, and others lie at distances greater than a certain $x^{**}(\epsilon)$, but none lie in between.
Those states at $x < x^*(\epsilon)$ belong to the same cluster as $\ket{\sigma}$, whereas those at $x > x^{**}(\epsilon)$ belong to different clusters.
We show this using the distance-resolved density of states, i.e., the number of states at distance $x$ with energy density $\epsilon'$.
It is exponential and written $e^{N g(x, \epsilon' | \epsilon)}$.
Using perturbation theory, we argue that the effective coupling between states is similarly exponential and written $e^{-N \gamma(x, \epsilon' | \epsilon)}$, which increases monotonically with $\Gamma$.
The two exponents $g(x, \epsilon' | \epsilon)$ and $\gamma(x, \epsilon' | \epsilon)$ are the central objects in our analysis.

Tunneling between clusters occurs only if there are states in different clusters which are resonant, i.e., states whose classical energies differ by less than the effective coupling between them.
The level spacing between states at distance $x$ and energy density $\epsilon$ is of order $e^{-N g(x, \epsilon | \epsilon)}$.
These resonate with the initial state $\ket{\sigma}$ if the effective coupling is larger than the level spacing at distance $x$.
The requirement for tunneling to occur is thus
\begin{equation} \label{eq:tunneling_condition}
\begin{gathered}
\max_{x \in [x^{**}(\epsilon), 1 - x^{**}(\epsilon)]} \big[ g(x, \epsilon | \epsilon) - \gamma(x, \epsilon | \epsilon) \big] > 0. \\
\textrm{\textbf{(Tunneling condition)}}
\end{gathered}
\end{equation}
The curve $\Gamma_{\textrm{tun}}(\epsilon)$ on which the left-hand side equals 0 defines the boundary between the trapped and tunneling phases.

As the initial state evolves in time, the amplitude on any non-resonant state remains exponentially small, of order $e^{-2N \gamma(x, \epsilon' | \epsilon)}$.
However, there are exponentially many such states.
If the \textit{total} amplitude on states with $\epsilon' \neq \epsilon$ is large, then those states will be observed at later times with high probability and the system excites into classical energy density $\epsilon'$.
The requirement for the classical energy density to be preserved is
\begin{equation} \label{eq:excitation_condition}
\begin{gathered}
\max_{x \in [0, 1]} \left[ \max_{\epsilon'} \big[ g(x, \epsilon' | \epsilon) - 2 \gamma(x, \epsilon' | \epsilon) \big] \right] < 0. \\
\textrm{\textbf{(Non-excitation condition)}}
\end{gathered}
\end{equation}
The curve $\Gamma_{\textrm{exc}}(\epsilon)$ on which the left-hand side equals 0 defines the boundary between the tunneling and excited phases.
There is an additional transition into the excited phase that coincides with the well-known thermodynamic transition into a quantum paramagnetic phase at large $\Gamma$.
We cannot detect this transition within perturbation theory, but we find clear evidence for it numerically.

Eqs.~\eqref{eq:tunneling_condition} and~\eqref{eq:excitation_condition} are necessary conditions for the quantum dynamics to succeed at energy matching.
If both are satisfied, then the time required for energy matching is simply the inverse tunneling rate between clusters.
We estimate the timescale $\tau$ using Fermi's golden rule with the effective coupling and distance-resolved density of states.
We find $\tau \sim e^{N r(\epsilon)}$ with
\begin{equation} \label{eq:tunneling_timescale}
\begin{gathered}
r(\epsilon) = \min_{x \in [x^{**}(\epsilon), 1 - x^{**}(\epsilon)]} \big[ 2 \gamma(x, \epsilon | \epsilon) - g(x, \epsilon | \epsilon) \big] . \\
\textrm{\textbf{(Tunneling timescale)}}
\end{gathered}
\end{equation}

The remainder of the paper is devoted to proving these results.
In Sec.~\ref{subsec:discussion_model}, we introduce the $p$-spin model $H_p$ which we use for our analysis.
We demonstrate the clustering of its low-energy states in Sec.~\ref{subsec:clustering}, and calculate the function $g(x, \epsilon' | \epsilon)$ in Sec.~\ref{subsec:franz_parisi}.
In Sec.~\ref{sec:large_p}, we consider the quantum dynamics of the full Hamiltonian $H$ for the simple and well-controlled Random Energy Model (the $p \rightarrow \infty$ limit of the $p$-spin model).
We develop the perturbation theory and calculate $\gamma(x, \epsilon' | \epsilon)$ in Sec.~\ref{subsec:perturbative_analysis}, compare the resulting tunneling rate to the Arrhenius rate for a classical Monte Carlo simulation in Sec.~\ref{subsec:thermal_comparison}, and use exact diagonalization of small systems to validate the perturbative results in Sec.~\ref{subsec:numerics}.
In Sec.~\ref{sec:finite_p}, we show that these results are robust in the more realistic mean-field models at large but finite $p$.
Finally, we conclude in Sec.~\ref{sec:conclusion}.

\section{The p-spin model} \label{sec:p_spin_model}

\subsection{Discussion of the model} \label{subsec:discussion_model}

We consider energy matching in the classical $p$-spin model, which consists of $N$ spin-1/2s with random all-to-all $p$-body interactions.
Although originally introduced as a mean-field model for spin glasses~\cite{Derrida1980Random,Gross1984Simplest,Gardner1985Spin}, the $p$-spin model has since received attention for its connections to structural glasses~\cite{Kirkpatrick1987pSpin,Biroli2012Random} and combinatorial optimization problems~\cite{Mezard2009,Bapst2013Quantum}.
Its theoretical versatility is due to the particularly simple Gaussian correlations between the energy levels.
It serves as an analytically tractable model of high-dimensional rugged energy landscapes.

The $p$-spin Hamiltonian is
\begin{equation} \label{eq:p_spin_hamiltonian}
H_p = \sum_{(i_1 \cdots i_p)} J_{i_1 \cdots i_p} \hat{\sigma}_{i_1}^z \cdots \hat{\sigma}_{i_p}^z.
\end{equation}
The sum is over all $p$-tuples of the $N$ spins.
Each coupling $J_{i_1 \cdots i_p}$ is an independent Gaussian random variable of mean 0 and variance $\frac{p!}{2N^{p-1}}$.
Denote a configuration of the spins by $\sigma$, where the $i$'th spin has value $\sigma_i$. From Eq.~\eqref{eq:p_spin_hamiltonian}, it follows that the energies are Gaussian-distributed with mean 0 and covariance matrix
\begin{equation} \label{eq:p_spin_cov_matrix}
\mathbb{E} \big[ H_p(\sigma) H_p(\sigma') \big] = \frac{N}{2} \Big( 1 - 2 x(\sigma, \sigma') \Big) ^p.
\end{equation}
Here and for the entirety of the paper, $\mathbb{E}$ denotes averages over the disordered couplings $J_{i_1 \cdots i_p}$.
$x(\sigma, \sigma')$ denotes the fractional Hamming distance between $\sigma$ and $\sigma'$ (Eq.~\eqref{eq:fractional_distance_def}).
Eq.~\eqref{eq:p_spin_cov_matrix} summarizes why the $p$-spin model is a useful testbed for disordered systems: the energy landscape is Gaussian-correlated with correlations that depend only on the distance in configuration space.
The parameter $p$ sets the strength of the correlations and is a useful parameter to vary.
In particular, as $p \rightarrow \infty$ the energy levels become independent~\cite{Derrida1980Random}.
When we consider quantum dynamics below, we use the $p \rightarrow \infty$ limit as a starting point and then argue that leading-in-$1/p$ corrections do not affect the qualitative picture.

\subsection{Clustering} \label{subsec:clustering}

The static and dynamical behavior of the $p$-spin model derives from the organization of its energy levels in configuration space.
Central to this organization is whether configurations at the same energy density are ``connected''.
We say that $\sigma$ and $\sigma'$ are connected (as $N \rightarrow \infty$) if one can be transformed into the other by a series of spin-flips, each flipping only $O(1)$ spins, while incurring only $O(1)$ changes in energy.
For a given $\sigma$, the set of $\sigma'$ to which it is connected defines a ``cluster''.
The motivation for these definitions is that, heuristically, stochastic dynamics such as Glauber or Metropolis quickly explores within a cluster but requires much longer times to transition between them.
In many physical and computational problems, including the $p$-spin model, the number and geometry of clusters transitions sharply at certain energy densities.

\begin{figure}[t]
\begin{center}
\includegraphics[width=1.0\columnwidth]{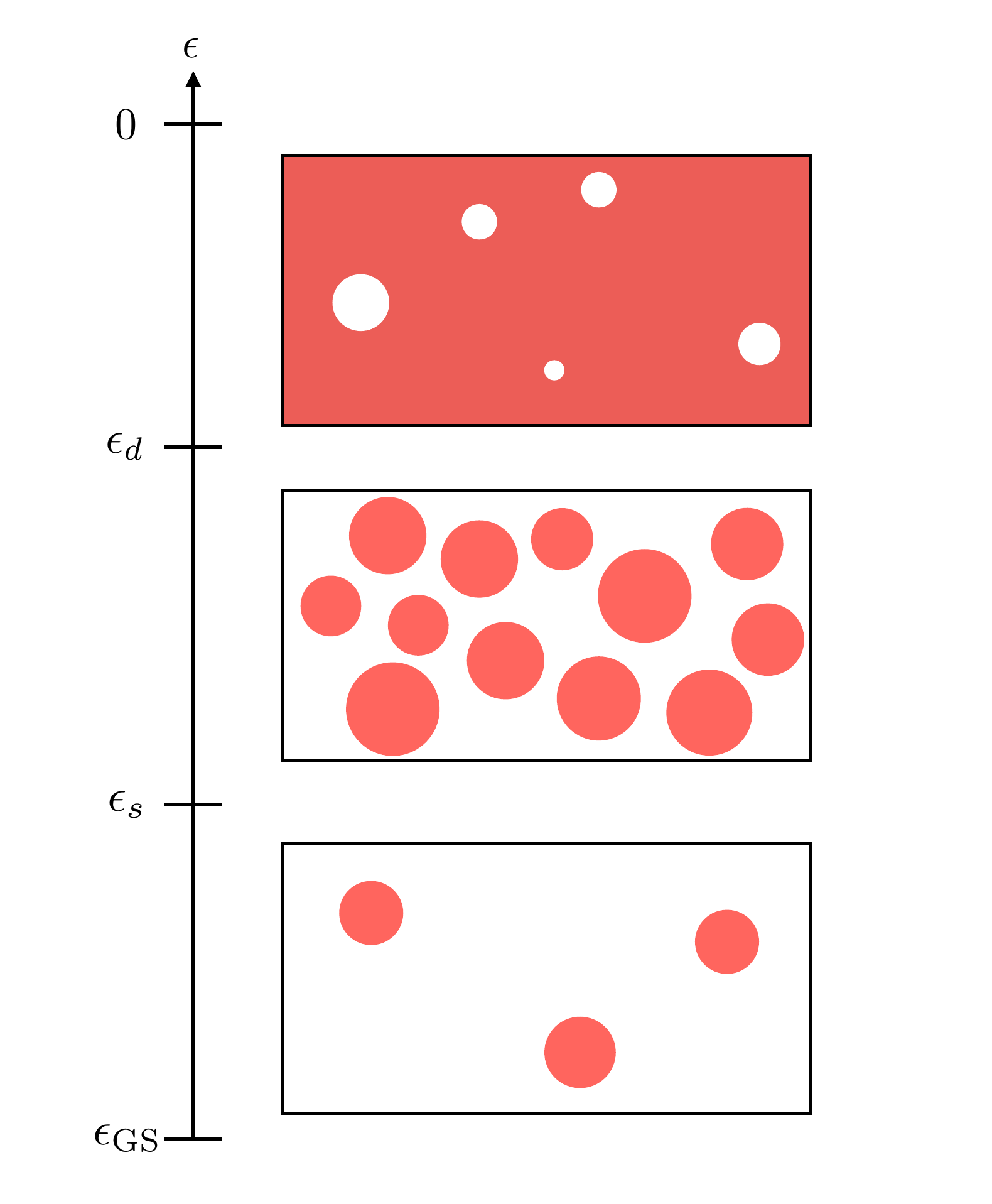}
\caption{Important transitions in the classical $p$-spin model. Each box represents the configuration space of $N$ spin-1/2s, and the red areas represent the regions of configuration space that contain states of energy density $\epsilon$. Top: $\epsilon_d < \epsilon < 0$, middle: $\epsilon_s < \epsilon < \epsilon_d$, bottom: $\epsilon_{\textrm{GS}} < \epsilon < \epsilon_s$.}
\label{fig:classical_transitions}
\end{center}
\end{figure}

\begin{figure}[t]
\begin{center}
\includegraphics[width=1.0\columnwidth]{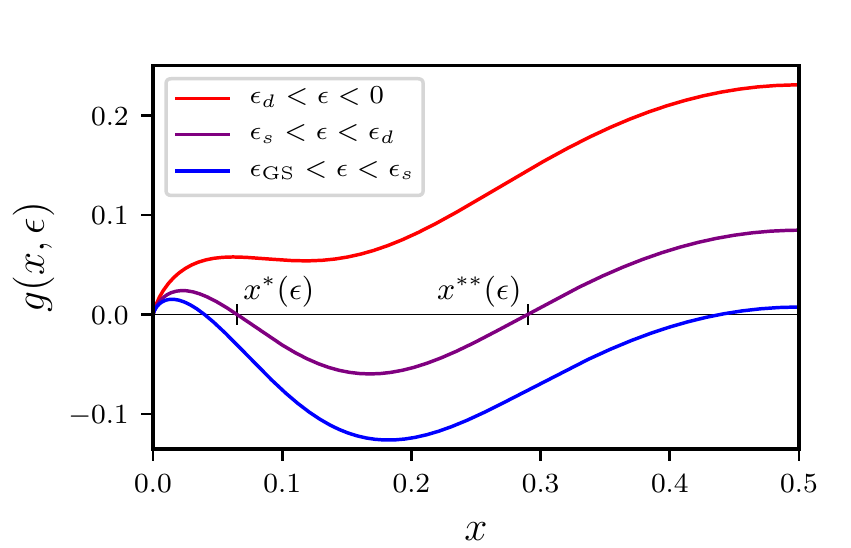}
\caption{Transitions in the $p$-spin model exhibited through the FPP, $g(x, \epsilon | \epsilon)$. Compare to Fig.~\ref{fig:classical_transitions}. Only $x < 1/2$ is shown, since (for even $p$) $g(x, \epsilon | \epsilon)$ is symmetric between $x \leftrightarrow 1 - x$. For the curves shown, we made the annealed approximation using $p = 6$ and $\epsilon = -0.68$ (red), $-0.78$ (purple), $-0.828$ (blue).}
\label{fig:entropy_curves}
\end{center}
\end{figure}

The relevant transitions for the present paper are sketched in Fig.~\ref{fig:classical_transitions}.
The center of the spectrum, corresponding to infinite temperature, is at energy density $\epsilon = 0$, and the bottom is at a finite energy density $\epsilon_{GS}$ ($< 0$).
In between, transitions occur at $\epsilon_d$ and $\epsilon_s$.
They are best understood in terms of two order parameters.
The Edwards-Anderson order parameter quantifies \textit{dynamical} ergodicity-breaking, e.g., in a Monte Carlo simulation run for time $t$:
\begin{equation} \label{eq:edwards_anderson_op}
q_{\textrm{EA}} \equiv \lim_{t \rightarrow \infty} \lim_{N \rightarrow \infty} \mathbb{E} \Big[ \Big< N^{-1} \sum_i \sigma_i (t) \sigma_i(0) \Big> \Big] .
\end{equation}
The angular brackets denote a thermal average over $\sigma(0)$ and an average over the randomness of the dynamics.
The order parameter that quantifies \textit{equilibrium} ergodicity-breaking is in terms of ``replicas'', i.e., copies of the system that are uncoupled from each other but have the same disorder realization:
\begin{equation} \label{eq:overlap_op}
q \equiv \lim_{N \rightarrow \infty} \mathbb{E} \Big[ \Big< \big| N^{-1} \sum_i \sigma_i^{\alpha} \sigma_i^{\beta} \big| \Big> \Big] .
\end{equation}
The superscripts $\alpha$ and $\beta$ denote different replicas and the angular brackets denote independent thermal averages over $\sigma^{\alpha}$ and $\sigma^{\beta}$.
Note that $N^{-1} \sum_i \sigma_i^{\alpha} \sigma_i^{\beta}$, the ``overlap'' between $\alpha$ and $\beta$, is simply $1 - 2x(\sigma^{\alpha}, \sigma^{\beta})$.
Two configurations chosen \textit{uniformly} out of all possible $2^N$ will have $q = 0$ ($x = 1/2$) with probability 1.
In a finite-temperature paramagnetic phase, $q$ remains at 0.
In any ordered phase, whether ferromagnetic or spin-glass, $q \neq 0$.

Now we turn to the relevant phases of the $p$-spin model.
\begin{itemize}
\item $\epsilon_d < \epsilon < 0$:
$q_{\textrm{EA}} = 0$, $q = 0$.
A randomly selected pair of states at such $\epsilon$ is connected with probability 1 (in the thermodynamic limit).
The corresponding cluster spans the configuration space, in the sense that the overlap between a randomly selected pair is 0 with probability 1 and stochastic dynamics equilibrates throughout the space.
\item $\epsilon_s < \epsilon < \epsilon_d$:
$q_{\textrm{EA}} \neq 0$, $q = 0$.
Typical states are no longer connected, and instead the number of clusters scales exponentially with $N$.
The timescale for transitioning between clusters is exponential in $N$.
In particular, it diverges in the thermodynamic limit, hence $q_{\textrm{EA}}$ is non-zero.
Nonetheless, the clusters are distributed throughout the configuration space.
A randomly selected pair of states belong to different clusters and the overlap is still 0.
\item $\epsilon_{\textrm{GS}} < \epsilon < \epsilon_s$: 
$q_{\textrm{EA}} \neq 0$, $q \neq 0$.
The number of clusters is $O(1)$ with respect to $N$.
A randomly selected pair of states has finite probability of belonging to the same cluster, which occupies only a small region of the configuration space.
This finite-probability event produces a non-zero average overlap, i.e., $q \neq 0$.
\end{itemize}
The transition at $\epsilon_d$ is called the ``dynamical'' or ``clustering'' transition, as it marks the energy density (or corresponding temperature) below which stochastic dynamics fails to equilibrate the system.
The transition at $\epsilon_s$ is called the ``static'' transition, as it is where the equilibrium order parameter becomes non-trivial.

We focus on the range $\epsilon_s < \epsilon < \epsilon_d$ in the present paper, for which the configurations are organized into exponentially many clusters.
The matching problem is solved if the system, initially prepared in one cluster, is found in a different cluster at a later time.

\subsection{The Franz-Parisi potential} \label{subsec:franz_parisi}

We study the geometry of clusters by computing the quantity defined in Eq.~\eqref{eq:microcanonical_fpp}.
It counts, for a fixed configuration $\sigma$ having energy density $\epsilon$, the number of configurations $\sigma'$ having energy density $\epsilon'$ which are separated by a distance $x$.
Precisely,
\begin{equation} \label{eq:microcanonical_fpp}
g(x, \epsilon' | \sigma, \epsilon) \equiv \frac{1}{N} \mathbb{E} \Big[ \ln{\textrm{Tr}_{\sigma'} \big[ \delta_{x, x(\sigma, \sigma')} } \, \delta \big( \epsilon' - \epsilon(\sigma') \big) \big] \Big] _{\epsilon(\sigma) = \epsilon}.
\end{equation}
The subscript to the disorder average indicates that we condition on having $\epsilon(\sigma) = \epsilon$.
The argument of the logarithm is the number of configurations at distance $x$ from $\sigma$ with energy density $\epsilon'$.
Thus $g(x, \epsilon' | \sigma, \epsilon)$ is the conditioned average entropy density at distance $x$.
Since the correlations between $\sigma$ and $\sigma'$ depend only on their separation, $g(x, \epsilon' | \sigma, \epsilon)$ depends on $\sigma$ only through $x$ and $\epsilon$.
Thus we will write $g(x, \epsilon' | \epsilon)$ throughout the paper.

$g(x, \epsilon' | \epsilon)$ is closely related to the Franz-Parisi potential (FPP), which is an important tool in the analysis of mean-field disordered systems~\cite{Kurchan1993Barriers,Franz1995Recipes,Franz1997Phase}.
Eq.~\eqref{eq:microcanonical_fpp} is essentially the FPP written in the microcanonical ensemble, as we describe in Appendix~\ref{appendix:fpp}.
We shall refer to $g(x, \epsilon' | \epsilon)$ as the FPP throughout the paper, but keep in mind that Eq.~\eqref{eq:microcanonical_fpp} is not the standard form in which it is presented.

Strictly speaking, one would need to use the replica trick to evaluate the FPP, as done in~\cite{Franz1995Recipes}.
However, the essential physics remains intact if we instead take an ``annealed'' average by switching the order of the logarithm and disorder average:
\begin{equation} \label{eq:annealed_fpp}
\begin{aligned}
g(x, \epsilon' | \epsilon) \approx & \; \frac{1}{N} \ln{ \mathbb{E} \Big[ \textrm{Tr}_{\sigma'} \big[ \delta_{x, x(\sigma, \sigma')} } \, \delta \big( \epsilon' - \epsilon(\sigma') \big) \big] \Big] _{\epsilon(\sigma) = \epsilon} \\
= & \; -x \ln{x} - (1 - x) \ln{(1 - x)} \\
& \qquad \qquad + \frac{1}{N} \ln{\mathbb{E} \Big[ \delta \big( \epsilon' - \epsilon(\sigma') \big) \Big] _{\epsilon(\sigma) = \epsilon}}.
\end{aligned}
\end{equation}
Note that $\mathbb{E} [ \delta ( \epsilon' - \epsilon(\sigma')) ] _{\epsilon(\sigma) = \epsilon}$ is the probability of $\epsilon(\sigma') = \epsilon'$ conditioned on $\epsilon(\sigma) = \epsilon$.
We present the calculation of this conditional distribution in Appendix~\ref{appendix:energy_correlations}.
The end result for the FPP is
\begin{equation} \label{eq:microcanonical_fpp_result}
\begin{aligned}
g(x, \epsilon' | \epsilon) \sim & \; -x \ln{x} - (1 - x) \ln{(1 - x)} \\
& \qquad \qquad \qquad - \frac{\big( \epsilon' - (1 - 2x)^p \epsilon \big) ^2}{1 - (1 - 2x)^{2p}}.
\end{aligned}
\end{equation}
Although Eq.~\eqref{eq:microcanonical_fpp_result} is only an annealed average, the inequality $\mathbb{E} \big[ \ln{( \cdot )} \big] \leq \ln{\mathbb{E} \big[ \cdot \big]}$ shows that it is a rigorous upper bound to the exact FPP.
In particular, if $g(x, \epsilon' | \epsilon) < 0$ then there are \textit{no} states having $\epsilon'$ at distance $x$ (with probability 1 in the thermodynamic limit)~\cite{Mezard2009}.

Define $g(x, \epsilon) \equiv g(x, \epsilon | \epsilon)$.
As a function of $x$, $g(x, \epsilon)$ demonstrates that low-lying energy levels are organized into clusters.
Fig.~\ref{fig:entropy_curves} gives representative examples.
Compare the shapes of $g(x, \epsilon)$ in Fig.~\ref{fig:entropy_curves} to the sketches in Fig.~\ref{fig:classical_transitions}.
\begin{itemize}
\item $\epsilon_d < \epsilon < 0$: 
$g(x, \epsilon) > 0$ for all $x$.
There are configurations that have the same energy density at all distances from the reference state $\sigma$.
This suggests that each configuration is connected to all others, forming a single cluster that spans the configuration space.
While it is not a proof, as the FPP distinguishes only the radial coordinate $x$ of configurations and not angular coordinates, dynamical calculations of the classical stochastic dynamics confirm that $q_{\textrm{EA}} = 0$ above $\epsilon_d$~\cite{Sompolinsky1981Dynamic,Kirkpatrick1987pSpin}.
\item $\epsilon_s < \epsilon < \epsilon_d$:
$g(x, \epsilon)$ is positive for $x$ less than a certain $x^*(\epsilon)$ or greater than a certain $x^{**}(\epsilon)$ (see Fig.~\ref{fig:entropy_curves}), but is negative in between.
This \textit{proves} that configurations below $\epsilon_d$ are organized into disjoint clusters.
No configurations at distances $x \in \big( x^*(\epsilon), x^{**}(\epsilon) \big)$ have energy density $\epsilon$, thus those at $x > x^{**}(\epsilon)$ cannot be connected to $\sigma$.
Furthermore, the maximum of $g(x, \epsilon)$ over $x > x^{**}(\epsilon)$ is greater than that over $x < x^*(\epsilon)$.
There are exponentially more configurations disconnected to $\sigma$ than connected, i.e., exponentially many clusters.
A randomly selected $\sigma'$ lies at distance 1/2 from $\sigma$.
\item $\epsilon_{\textrm{GS}} < \epsilon < \epsilon_s$:
The maximum of $g(x, \epsilon)$ over $x > x^{**}(\epsilon)$ is now \textit{less} than that over $x < x^*(\epsilon)$.
Interpreting this result literally, one would say that most configurations belong to a single cluster of linear size $x^*(\epsilon)$.
A randomly selected $\sigma'$ lies within that distance.
\end{itemize}

Keep in mind that since we estimated $g(x, \epsilon)$ through an annealed average, Eq.~\eqref{eq:microcanonical_fpp_result} gives only approximate values for the quantities defined above ($\epsilon_d$, $\epsilon_s$, $x^*(\epsilon)$, etc.).
In particular, the interpretation that below $\epsilon_s$ most states belong to a single cluster is too naive: the number of clusters is $O(1)$ but larger than 1, and it depends on $\epsilon$~\cite{Mezard2009}.
However, $\epsilon_d$ as estimated from Eq.~\eqref{eq:microcanonical_fpp_result} is an exact lower bound on the location of the clustering transition.
Below we shall need to compute the energy barriers between clusters, and here as well the annealed estimate gives exact lower bounds.

\section{Quantum dynamics in the large-p limit} \label{sec:large_p}

Here we describe the performance of quantum dynamics in tunneling between the clusters of the $p$-spin model, specifically in the $p \rightarrow \infty$ limit.
Our main results are the tunneling and non-excitation conditions, Eqs.~\eqref{eq:tunneling_condition} and~\eqref{eq:excitation_condition} respectively, both of which are necessary conditions for the dynamics to succeed in energy-matching.
They express that the tunneling amplitudes must be large enough to hybridize states between clusters, but not so large that the system is excited out of clusters.
Even if both requirements are satisfied, the tunneling rate is exponentially slow in system size, with the exponent given by Eq.~\eqref{eq:tunneling_timescale}.

The Hamiltonian that we consider is
\begin{equation} \label{eq:quantum_p_spin_hamiltonian}
H = H_p - \Gamma \sum_{i} \hat{\sigma}_i^x \equiv H_p + V,
\end{equation}
with $H_p$ as in Eq.~\eqref{eq:p_spin_hamiltonian}, which we refer to as the ``classical'' energy.
The second term, a uniform transverse field, causes spin-flips.
In the ($\sigma^z$) configuration space, it acts as a hopping term while $H_p$ acts as a potential.
Thus Eq.~\eqref{eq:quantum_p_spin_hamiltonian} can be interpreted as an Anderson problem~\cite{Anderson1958Absence,Thouless1974Electrons} in the many-body configuration space.
The $p \rightarrow \infty$ limit corresponds to an uncorrelated potential and small clusters, and is simplest to study for reasons which we describe below.
The bottom panel of Fig.~\ref{fig:REM_phase_diagram} presents the dynamical phase diagram in this limit.
In the ``trapped'' phase, the system does not tunnel between clusters, even on exponentially long timescales.
In the ``excited'' phase, the system is excited to higher classical energy densities.
Only in the ``tunneling'' phase does the system tunnel between clusters and succeed in energy matching.

We demonstrate these results using perturbation theory in Sec.~\ref{subsec:perturbative_analysis}.
In Sec.~\ref{subsec:thermal_comparison}, we show that the tunneling rates thus obtained, although exponentially slow in system size, provide exponential speed-up over classical Monte Carlo simulations.
We numerically validate the dynamical phase diagram in Sec.~\ref{subsec:numerics}.

\subsection{Perturbative analysis} \label{subsec:perturbative_analysis}

In the $p \rightarrow \infty$ limit, Eq.~\eqref{eq:p_spin_cov_matrix} for the correlation between the classical energies becomes
\begin{equation} \label{eq:REM_cov_matrix}
\mathbb{E} \left[ H_p(\sigma) H_p(\sigma') \right] \rightarrow \frac{N}{2} \delta_{0, x(\sigma, \sigma')} = \frac{N}{2} \delta_{\sigma, \sigma'}.
\end{equation}
The classical energies are independent and distributed as
\begin{equation} \label{eq:REM_energy_distribution}
P_1 (\epsilon) = \sqrt{\frac{N}{\pi}} e^{-N \epsilon^2},
\end{equation}
and the FPP is
\begin{equation} \label{eq:REM_microcanonical_fpp}
g(x, \epsilon' | \epsilon) = -x \ln{x} - (1 - x) \ln{(1 - x)} - \big( \epsilon' \big) ^2.
\end{equation}
The $p \rightarrow \infty$ model is referred to as the ``Random Energy Model'' (REM)~\cite{Derrida1980Random}.
It is the simplest context in which to study tunneling between clusters because the clusters have no internal structure: $g(x, \epsilon) < 0$ for \textit{all} $x$ less than a certain $x^{**}(\epsilon)$, i.e., each ``cluster'' has size 0 and in fact consists of a single configuration.
$x^{**}(\epsilon)$ is the minimum distance between any configurations having energy density $\epsilon$.

\begin{figure}[t]
\begin{center}
\includegraphics[width=1.0\columnwidth]{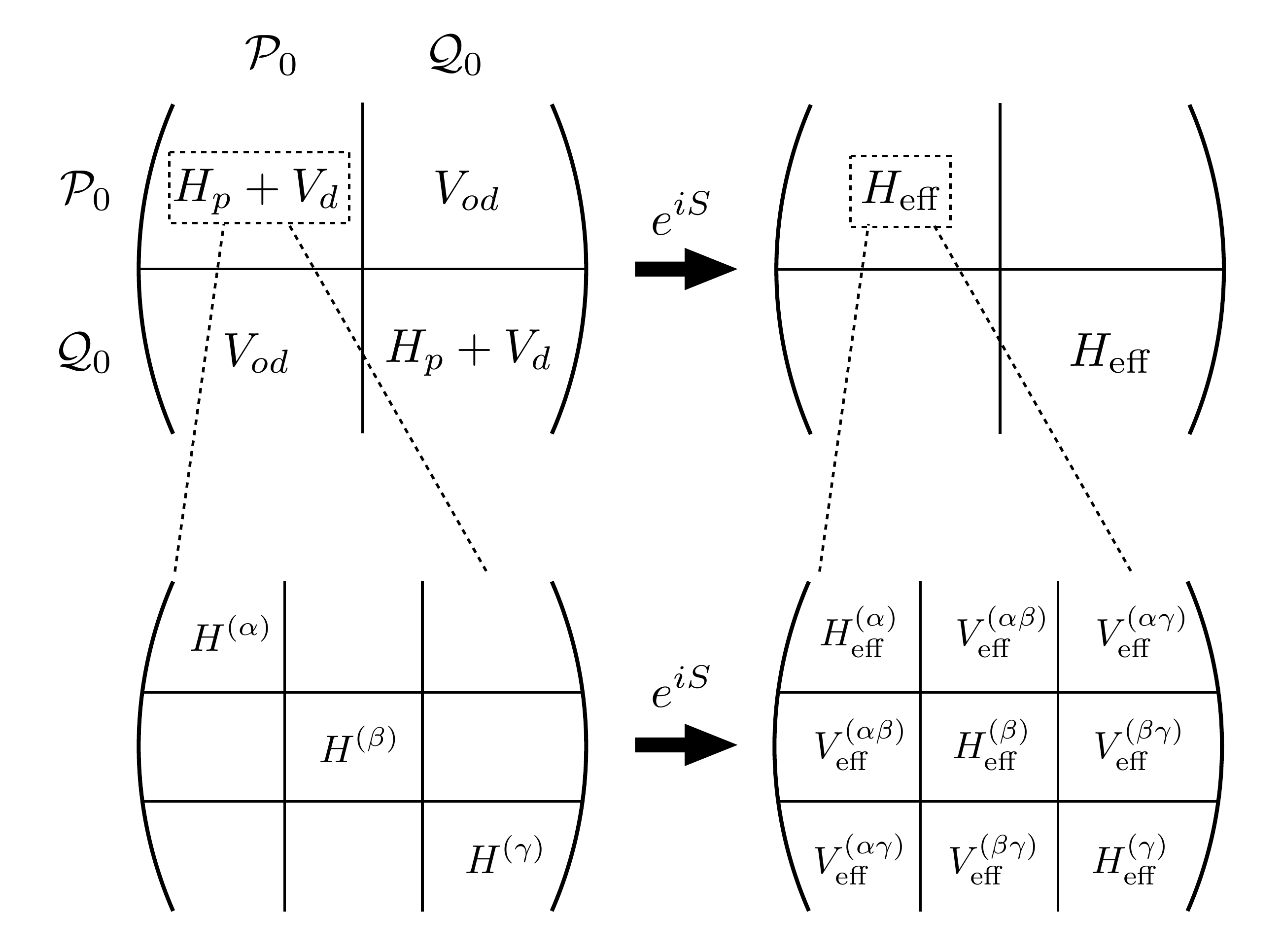}
\caption{The effect of the Schrieffer-Wolff transformation. The left side shows $H$, the right side shows $H_{\textrm{eff}}$. The top shows the full Hamiltonian, broken into $\mathcal{P}_0$ and $\mathcal{Q}_0$ subspaces. The transverse field operator $V$ is broken into a block-diagonal part $V_d$ and a block-off-diagonal part $V_{od}$. The bottom is a schematic of the structure within $\mathcal{P}_0$. The superscripts refer to different clusters.}
\label{fig:Schrieffer_Wolff}
\end{center}
\end{figure}

To study tunneling between clusters, we formally use the Schrieffer-Wolff transformation~\cite{Bravyi2011Schrieffer} together with the forward-scattering approximation~\cite{Pietracaprina2016Forward,Baldwin2016Many}.
First, we discuss the Schrieffer-Wolff transformation.
Let $\mathcal{P}_0$ denote the subspace spanned by $\sigma^z$ configurations having classical energy density $\epsilon$, and let $\mathcal{Q}_0$ denote the orthogonal subspace.
We take $\epsilon < \epsilon_d$, so that the configurations within $\mathcal{P}_0$ are organized into clusters.
Note that $H$ couples $\mathcal{P}_0$ and $\mathcal{Q}_0$ through the transverse field, yet it does not directly couple configurations within different clusters, as multiple spin-flips would be required.
The Schrieffer-Wolff transformation is performed by a unitary operator $e^{iS}$ such that $e^{iS} H e^{-iS} \equiv H_{\textrm{eff}}$ \textit{does not} couple $\mathcal{P}_0$ and $\mathcal{Q}_0$.
In return, $H_{\textrm{eff}}$ does have a direct coupling between configurations within different clusters, denoted $V_{\textrm{eff}}$.
The situation is illustrated in Fig.~\ref{fig:Schrieffer_Wolff}.
We denote by $H^{(\alpha)}$ the projection of $H$ into cluster $\alpha$, and similarly for $H_{\textrm{eff}}^{(\alpha)}$ and $V_{\textrm{eff}}^{(\alpha \beta)}$.
Since
\begin{equation} \label{eq:time_evolution_transformation}
\braket{\sigma' | e^{-i H t} | \sigma} = \bra{\sigma'} e^{-iS} e^{-i H_{\textrm{eff}} t} e^{iS} \ket{\sigma} ,
\end{equation}
the time evolution of $\ket{\sigma}$ into $\ket{\sigma'}$ under $H$ is equivalent to the evolution of $e^{iS} \ket{\sigma}$ into $e^{iS} \ket{\sigma'}$ under $H_{\textrm{eff}}$.

\begin{figure}[t]
\begin{center}
\includegraphics[width=1.0\columnwidth]{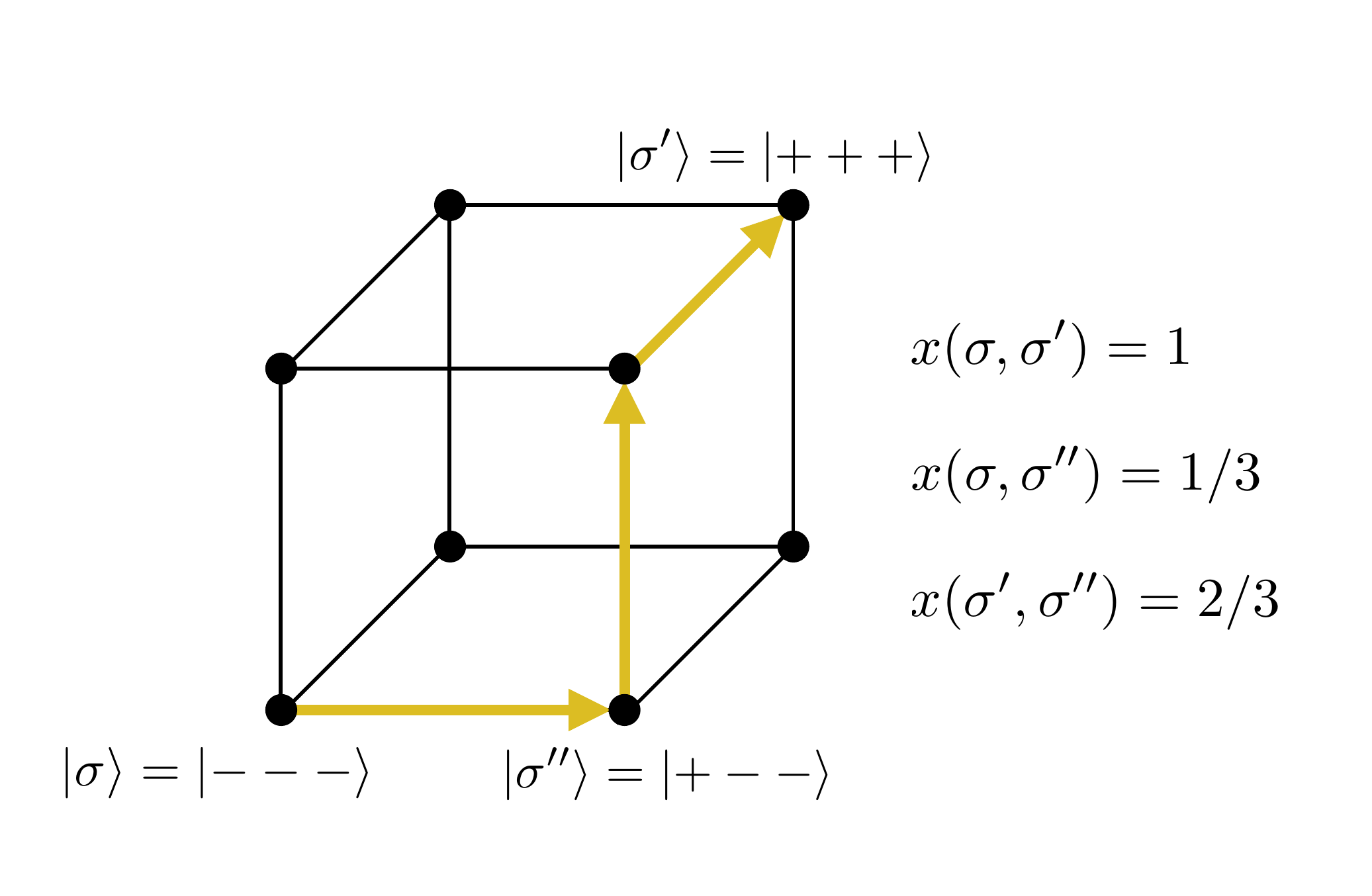}
\caption{The configuration space (black vertices) for $N = 3$. The gold arrows form a directed path from $\ket{\sigma}$ to $\ket{\sigma'}$. $\ket{\sigma''}$ is an intermediate state. Distances between configurations are shown on the right.}
\label{fig:hypercube_example}
\end{center}
\end{figure}

The method by which one calculates the generator $S$ is known in the literature~\cite{Bravyi2011Schrieffer}.
We quote it in Appendix~\ref{appendix:schrieffer_wolff}, where we also detail the forward-scattering approximation (FSA).
The FSA approximates $\braket{\sigma' | V_{\textrm{eff}} | \sigma}$ (where $\epsilon(\sigma) = \epsilon(\sigma') = \epsilon$) by its lowest-order terms in the transverse field $\Gamma$.
As reviewed in Appendix~\ref{appendix:schrieffer_wolff}, they correspond to directed ``paths'' in the configuration space, i.e., sequences of spin-flips that transform $\sigma$ into $\sigma'$:
\begin{equation} \label{eq:forward_scattering_expression}
\braket{\sigma' | V_{\textrm{eff}} | \sigma} \sim \sum_P \Gamma \prod_{\sigma'' \in P} \frac{\Gamma}{N \big( \epsilon - \epsilon(\sigma'') \big) }.
\end{equation}
The sum is over the $\big( N x(\sigma, \sigma') \big) !$ sequences of spin-flips, and the product is over the intermediate configurations along each sequence. Fig.~\ref{fig:hypercube_example} gives an example of such a path.

Each $\epsilon(\sigma'')$ in Eq.~\eqref{eq:forward_scattering_expression} is an independent random variable of mean 0.
If we replace every $\epsilon(\sigma'')$ by $\mathbb{E} \big[ \epsilon(\sigma'') \big]$, the effective coupling takes a simple form:
\begin{equation} \label{eq:REM_forward_scattering_evaluation}
\big| \braket{\sigma' | V_{\textrm{eff}} | \sigma} \big| \sim (Nx)! \left( \frac{\Gamma}{N |\epsilon|} \right) ^{Nx} \sim e^{-N \gamma(x, \epsilon)},
\end{equation}
where
\begin{equation} \label{eq:REM_effective_coupling_exponent}
\gamma(x, \epsilon) \equiv -x \ln{\frac{x \Gamma}{e |\epsilon|}}.
\end{equation}

It turns out that the distribution of $\big| \braket{\sigma' | V_{\textrm{eff}} | \sigma} \big|$ over realizations concentrates around this value.
Write
\begin{equation} \label{eq:denominators_to_exponent}
\prod_{\sigma'' \in P} \frac{1}{\big| \epsilon - \epsilon(\sigma'') \big|} = e^{- \sum \ln{\big| \epsilon - \epsilon(\sigma'') \big|}}.
\end{equation}
In order for Eq.~\eqref{eq:denominators_to_exponent} to scale as anything other than $\exp{\big( -Nx \ln{|\epsilon|} \big) }$, it must be that $O(N)$ of the $\sigma''$ along path $P$ have $\epsilon(\sigma'') \neq 0$.
The probability of such an event scales as $e^{-c N^2}$, with $c \sim O(1)$.
Thus the expected number of paths which contribute anything other than $\exp{\big( -Nx \ln{|\epsilon|} \big) }$, denoted $\mathcal{N}_{\textrm{fluc}}$, is
\begin{equation} \label{eq:expected_fluctuating_paths}
\mathbb{E} [ \mathcal{N}_{\textrm{fluc}} ] = (Nx)! \, e^{-c N^2} \rightarrow 0,
\end{equation}
i.e., the expected number of atypical paths vanishes as $N \rightarrow \infty$.
Eq.~\eqref{eq:REM_forward_scattering_evaluation} thus gives the effective coupling between clusters in the $p \rightarrow \infty$ limit.

We next use time-dependent perturbation theory to calculate the transition amplitude from $\ket{\sigma}$ to $\ket{\sigma'}$ given in Eq.~\eqref{eq:time_evolution_transformation}.
We can take $e^{iS} \ket{\sigma} \sim \ket{\sigma}$ and $e^{iS} \ket{\sigma'} \sim \ket{\sigma'}$ without affecting the lowest-order terms (see Appendix~\ref{appendix:schrieffer_wolff}).
Furthermore, we neglect ``self-energy'' corrections that would modify the diagonal elements of $H_{\textrm{eff}}$: the corrections to the energy \textit{densities} are only $O(1/N)$~\cite{Jorg2008Simple}, and the classical energies are distributed randomly regardless.
Then the standard derivation~\cite{Sakurai2011} gives
\begin{widetext}
\begin{equation} \label{eq:transition_prob}
\big| \braket{\sigma' | e^{-i H t} | \sigma} \big|^2 = \frac{4 \big| \braket{\sigma' | V_{\textrm{eff}} | \sigma} \big| ^2}{N^2 \big( \epsilon(\sigma') - \epsilon(\sigma) \big) ^2} \sin^2{\left( \frac{N \big( \epsilon(\sigma') - \epsilon(\sigma) \big) t}{2} \right) }.
\end{equation}
\end{widetext}

We are interested in $\epsilon(\sigma') \sim \epsilon(\sigma) \sim \epsilon$, but it is important that we do not set $\epsilon(\sigma')$ exactly equal to $\epsilon(\sigma)$.
The smallest $\big| \epsilon(\sigma') - \epsilon(\sigma) \big|$ over all $\sigma'$ is typically of the order of level spacing, which may be either smaller or larger than $\braket{\sigma' | V_{\textrm{eff}} | \sigma}$ since both scale exponentially with $N$.
More precisely, the level spacing \textit{among $\ket{\sigma'}$ at distance $x$} scales as $e^{-N g(x, \epsilon)}$, with $g(x, \epsilon)$ given by Eq.~\eqref{eq:REM_microcanonical_fpp} (setting $\epsilon' = \epsilon$).
If $g(x, \epsilon) < \gamma(x, \epsilon)$, then Eq.~\eqref{eq:transition_prob} is exponentially small for every final state at distance $x$.
Furthermore, the total weight on all states at distance $x$, obtained by summing Eq.~\eqref{eq:transition_prob} over all $\ket{\sigma'}$ with $\epsilon(\sigma') \sim \epsilon(\sigma)$ and $x(\sigma, \sigma') = x$, is exponentially small: those $\ket{\sigma'}$ for which $\epsilon(\sigma') - \epsilon(\sigma) \sim e^{-N d}$ give a total contribution $e^{-2N \gamma + 2Nd} \cdot e^{Ng - Nd} = e^{-2N \gamma + Ng + Nd}$, which increases with $d$ until the smallest possible spacing at $d = g$.
Thus the probability of transitioning to any state at distance $x$ vanishes in the thermodynamic limit, for all times $t$.
Since clusters lie at distances $x \in \big[ x^{**}(\epsilon), 1 - x^{**}(\epsilon) \big]$, if $g(x, \epsilon) < \gamma(x, \epsilon)$ for all such $x$ then the system never transitions into a different cluster, even on exponentially long timescales.
This gives the tunneling condition presented in the introduction as a necessary condition for quantum dynamics to succeed in energy matching:
\begin{equation}
\begin{gathered}
\max_{x \in [x^{**}(\epsilon), 1 - x^{**}(\epsilon)]} \big[ g(x, \epsilon) - \gamma(x, \epsilon) \big] > 0. \tag{\ref{eq:tunneling_condition}} \\
\textrm{\textbf{(Tunneling condition)}}
\end{gathered}
\end{equation}
Using Eqs.~\eqref{eq:REM_microcanonical_fpp} and~\eqref{eq:REM_effective_coupling_exponent}, the tunneling condition becomes
\begin{equation} \label{eq:REM_tunneling_condition}
\max_{x \in [x^{**}(\epsilon), 1 - x^{**}(\epsilon)]} \left[ x \ln{\frac{\Gamma}{e |\epsilon|}} - (1 - x) \ln{(1 - x)} - \epsilon^2 \right] > 0,
\end{equation}
where $x^{**}(\epsilon)$ solves $-x \ln{x} - (1 - x) \ln{(1 - x)} = \epsilon^2$.
The curve $\Gamma_{\textrm{tun}}(\epsilon)$ on which the left-hand side is 0 constitutes a sharp dynamical phase boundary, separating phases in which the system does and does not tunnel between clusters.

Note that the tunneling transition also manifests in properties of the many-body \textit{eigenstates}, as discussed in Ref.~\cite{Laumann2014Many}.
At $\Gamma < \Gamma_{\textrm{tun}}(\epsilon)$, the eigenstates are only slightly perturbed from the classical states $\ket{\sigma}$, whereas at $\Gamma > \Gamma_{\textrm{tun}}(\epsilon)$, the eigenstates are superposed from all classical states at energy density $\epsilon$.
This picture was confirmed by extensive numerics in Ref.~\cite{Baldwin2016Many}.
The use of perturbation theory and the FSA were also validated, by evaluating Eq.~\eqref{eq:forward_scattering_expression} \textit{numerically} for instances of finite-size systems.
It was found that the location of the tunneling transition predicted by the numerical FSA agrees very well with the location found by exact diagonalization.

If Eq.~\eqref{eq:tunneling_condition} is satisifed, then Fermi's golden rule gives the \textit{rate} at which the system tunnels between clusters.
To ensure that we handle the exponentially small scales in the level spacing and effective coupling correctly, we present a derivation here.
Consider $\ket{\sigma'}$ at distance $x$ and fixed time $t$.
If $N \big| \epsilon(\sigma') - \epsilon(\sigma) \big| \ll t^{-1}$ then $\big| \braket{\sigma' | e^{-i H t} | \sigma} \big|^2$ behaves as $e^{-2N \gamma(x, \epsilon)} t^2$, whereas if $N \big| \epsilon(\sigma') - \epsilon(\sigma) \big| \gg t^{-1}$ then $\big| \braket{\sigma' | e^{-i H t} | \sigma} \big|^2$ has already reached its maximum value and begun oscillating.
Thus the portion of the weight at distance $x$ which is growing with time, denoted $P_{\textrm{tun}}(x, t)$, is obtained by summing Eq.~\eqref{eq:transition_prob} over $\ket{\sigma'}$ with $\big| \epsilon(\sigma') - \epsilon(\sigma) \big| \lesssim (Nt)^{-1}$.
To exponential order,
\begin{equation} \label{eq:fermi_golden_rule}
\begin{aligned}
P_{\textrm{tun}}(x, t) \sim & \, \left( e^{-2N \gamma(x, \epsilon)} t^2 \right) \left( e^{Ng(x, \epsilon)} t^{-1} \right) \\
= & \; e^{-N \big( 2 \gamma(x, \epsilon) - g(x, \epsilon) \big)} t,
\end{aligned}
\end{equation}
from which the tunneling rate is apparent.
The rate $\tau^{-1}$ at which the system exits its initial cluster is given by integrating over $x$, which since $P_{\textrm{tun}}(x, t)$ scales exponentially gives $\tau^{-1} \sim e^{-N r(\epsilon)}$ with $r(\epsilon)$ as stated in the introduction:
\begin{equation}
\begin{gathered}
r(\epsilon) = \min_{x \in [x^{**}(\epsilon), 1 - x^{**}(\epsilon)]} \big[ 2 \gamma(x, \epsilon) - g(x, \epsilon) \big] . \tag{\ref{eq:tunneling_timescale}} \\
\textrm{\textbf{(Tunneling timescale)}}
\end{gathered}
\end{equation}

A sufficiently strong transverse field is required to satisfy Eq.~\eqref{eq:tunneling_condition}, yet the perturbation theory breaks down if the transverse field is too strong.
$e^{iS} \ket{\sigma}$ has weight not only in the $\mathcal{P}_0$ subspace but also in $\mathcal{Q}_0$, and these two components of $e^{iS} \ket{\sigma}$ evolve separately over time (see Fig.~\ref{fig:Schrieffer_Wolff}).
In order for perturbation theory to be valid, the total weight in $\mathcal{Q}_0$ should be small.
Let $\ket{\sigma'}$ now be a state in $\mathcal{Q}_0$, so that $\epsilon(\sigma') \equiv \epsilon' \neq \epsilon$.
Within the same approximations as above, to lowest non-zero order
\begin{equation} \label{eq:outside_amplitude_expression}
\braket{\sigma' | e^{iS} | \sigma} \sim \braket{\sigma' | iS | \sigma} \sim \sum_P \prod_{\sigma'' \in P} \frac{\Gamma}{N \big( \epsilon - \epsilon(\sigma'') \big) }.
\end{equation}
The sum is again over the direct paths from $\ket{\sigma}$ to $\ket{\sigma'}$, and the product is again over intermediate states $\ket{\sigma''}$.

As before, we can take $\epsilon(\sigma'') \rightarrow \mathbb{E} \big[ \epsilon(\sigma'') \big] = 0$.
Eq.~\eqref{eq:outside_amplitude_expression} becomes
\begin{equation} \label{eq:outside_amplitude_evaluation}
\big| \braket{\sigma' | e^{iS} | \sigma} \big| \sim \left( \frac{x \Gamma}{e |\epsilon|} \right) ^{Nx} \equiv e^{-N \gamma(x, \epsilon' | \epsilon)}.
\end{equation}
Even though $\gamma(x, \epsilon' | \epsilon)$ does not depend on $\epsilon'$ (and in fact $\gamma(x, \epsilon' | \epsilon) = \gamma(x, \epsilon)$ from Eq.~\eqref{eq:REM_effective_coupling_exponent}), we keep mention of it in our notation.
The lack of $\epsilon'$-dependence is due to the lack of correlations in the $p \rightarrow \infty$ limit.
As soon as one includes finite-$p$ corrections, as in Sec.~\ref{sec:finite_p}, $\gamma(x, \epsilon' | \epsilon)$ depends on $\epsilon'$.

The total weight in the $\mathcal{Q}_0$ subspace is
\begin{equation} \label{eq:total_outside_weight}
\begin{aligned}
& \int_0^1 \textrm{d}x \int \textrm{d}\epsilon' \, \big| \braket{\sigma' | e^{iS} | \sigma} \big| ^2 e^{N g(x, \epsilon' | \epsilon)} \\
& \qquad \qquad \sim \int_0^1 \textrm{d}x \int \textrm{d}\epsilon' \, e^{N \big( g(x, \epsilon' | \epsilon) - 2 \gamma(x, \epsilon' | \epsilon) \big)}.
\end{aligned}
\end{equation}
The result must be exponentially small if perturbation theory is to be valid.
This is the non-excitation condition given in the introduction:
\begin{equation}
\begin{gathered}
\max_{x \in [0, 1]} \left[ \max_{\epsilon'} \big[ g(x, \epsilon' | \epsilon) - 2 \gamma(x, \epsilon' | \epsilon) \big] \right] < 0. \tag{\ref{eq:excitation_condition}} \\
\textrm{\textbf{(Non-excitation condition)}}
\end{gathered}
\end{equation}
For the REM, the maximization over $\epsilon'$ is trivial and we have the requirement
\begin{equation} \label{eq:REM_excitation_condition}
\max_{x \in [0, 1]} \left[ 2x \ln{\frac{\Gamma}{e |\epsilon|}} + x \ln{x} - (1 - x) \ln{(1 - x)} \right] < 0.
\end{equation}
Strictly speaking, violation of Eq.~\eqref{eq:excitation_condition} only means that the perturbation theory is inconsistent.
However, it has an immediate physical interpretation: the system excites out of its initial cluster and into higher classical energy densities.
This corresponds to another failure mechanism for energy-matching, as a measurement of the system will not yield the desired energy density.

\begin{figure}[t]
\begin{center}
\includegraphics[width=1.0\columnwidth]{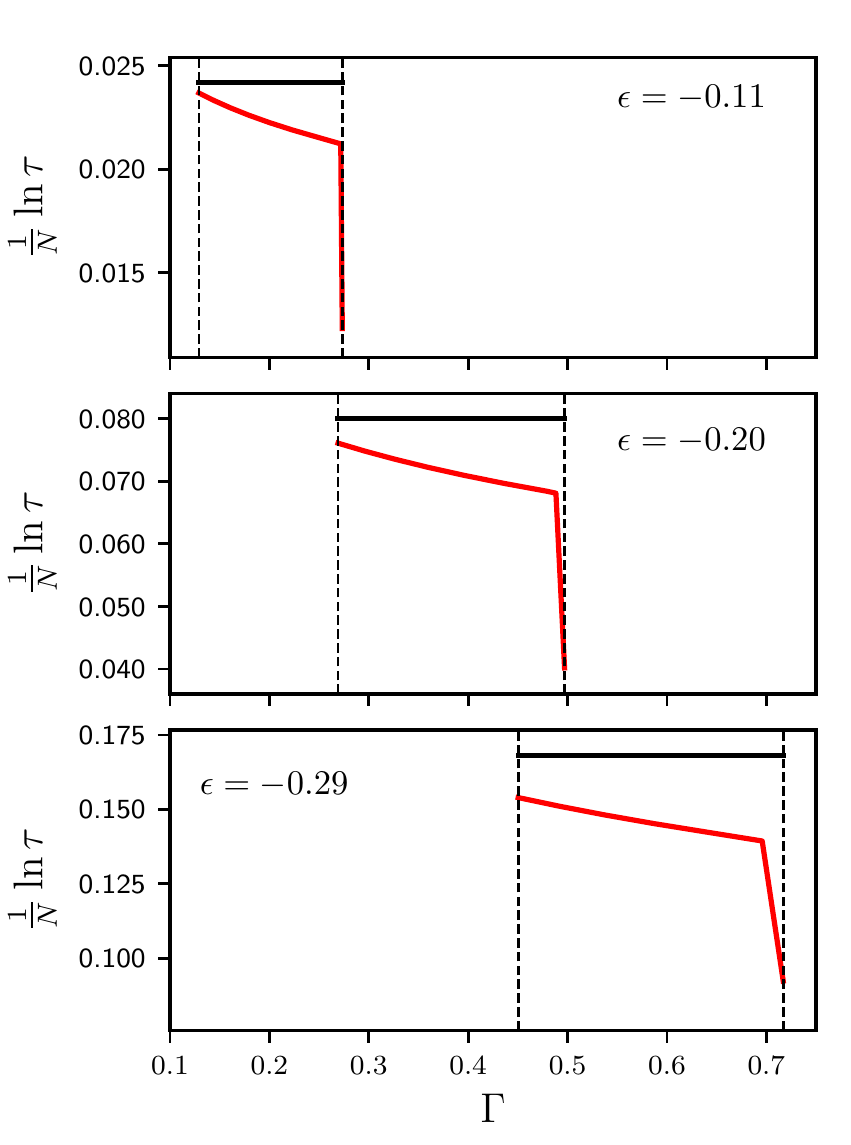}
\caption{Timescales for energy matching in the REM, as a function of $\Gamma$ at the indicated $\epsilon$. Red is $\tau_{\textrm{q}}$, black is $\tau_{\textrm{th}}$. In each panel, the left dashed line is the location of the tunneling transition and the right dashed line is the location of the excitation transition. $\tau_{\textrm{q}}$ is meaningful only in between the two. The non-analyticity in $\tau_{\textrm{q}}$ comes from a change in the argmin of Eq.~\eqref{eq:REM_tunneling_exponent}.}
\label{fig:tunneling_rates}
\end{center}
\end{figure}

Thus we have obtained two necessary conditions for quantum dynamics to succeed in energy-matching.
If Eq.~\eqref{eq:tunneling_condition} is violated, the system never escapes its initial cluster.
If Eq.~\eqref{eq:excitation_condition} is violated, the system excites out of clusters and does not return.
Generically, satisfying both conditions requires that the transverse field be neither too strong nor too weak, and this regime may be narrow or non-existent depending on $\epsilon$ and the model in question.
Indeed, Fig.~\ref{fig:REM_phase_diagram} shows that one cannot satisfy both conditions at low $\epsilon$ in the REM.

Ultimately, the Schrieffer-Wolff transformation and the FSA are uncontrolled approximations in this context.
In Sec.~\ref{subsec:numerics}, we therefore study the Random Energy Model numerically via exact diagonalization.
We find clear evidence for the tunneling transition, although finite-size effects prevent us from confirming its specific functional form.
Finite-size effects also prevent us from unambiguously identifying the excitation transition predicted by Eq.~\eqref{eq:excitation_condition}.
However, we do find that the first-order \textit{thermodynamic} transition into a quantum paramagnetic phase~\cite{Goldschmidt1990Solvable} is relevant for the Hamiltonian dynamics.
It is itself an excitation transition in which the system excites to classical energy density 0, and is another breakdown of energy-matching.
As the transition is first order, the perturbative expansion in $\Gamma$ does not locate it.
Instead, the phase boundary was calculated using the replica method in Ref.~\cite{Goldschmidt1990Solvable}.
We include it in Fig.~\ref{fig:REM_phase_diagram}, which highlights the limitations of quantum dynamics in energy-matching.

\subsection{Comparison to thermal activation rate} \label{subsec:thermal_comparison}

The rate of tunneling between clusters, when it occurs at all, is always exponentially slow in system size.
This is clear from the expressions for the rate (Eq.~\eqref{eq:tunneling_timescale}) and the non-excitation condition (Eq.~\eqref{eq:excitation_condition}): if Eq.~\eqref{eq:excitation_condition} is satisfied, then $r(\epsilon)$ in Eq.~\eqref{eq:tunneling_timescale} is necessarily positive since it optimizes over fewer variables.
Thus quantum dynamics cannot succeed at energy matching in polynomial time.
However, it may nevertheless be exponentially \textit{faster} than simple classical algorithms.
We now show this for the REM, by comparing the tunneling rate found above to the equilibration timescale for stochastic (e.g., Monte Carlo) dynamics.

Stochastic dynamics can be used for energy matching as follows: from a starting configuration $\sigma$, run a Monte Carlo simulation at the temperature $T$ corresponding to $\epsilon(\sigma)$ by Legendre transform.
Since the temperature is properly chosen, one will observe a configuration having the same energy density at later times with high probability.
However, the later configuration will belong to the same cluster unless one waits long enough for the system to be thermally activated over the energy barriers that separate clusters.
This activation timescale is thus the time required for Monte Carlo dynamics to succeed in energy matching.

It is straightforward to calculate the activation timescale in the REM.
The $N$ neighboring configurations to a given $\sigma$ all have energy density 0 with high probability, as follows from Eq.~\eqref{eq:REM_energy_distribution}.
Thus in a Monte Carlo simulation with single-spin update rules (e.g., Metropolis), the simulation time required to leave state $\sigma$ is $e^{N \beta |\epsilon(\sigma)|}$.
The thermodynamic entropy density of the REM is
\begin{equation} \label{eq:REM_entropy}
s(\epsilon) = \ln{2} - \epsilon^2,
\end{equation}
see Ref.~\cite{Mezard2009}, from which it follows that $\beta(\epsilon) = -2 \epsilon$.
The activation timescale in the REM starting from energy density $\epsilon$ is therefore
\begin{equation} \label{eq:activation_time}
\tau_{\textrm{th}} \sim e^{2N \epsilon^2}.
\end{equation}

The tunneling timescale is
\begin{equation} \label{eq:tunneling_time}
\tau_{\textrm{q}} \sim e^{N r(\epsilon)},
\end{equation}
with $r(\epsilon)$ given by Eq.~\eqref{eq:tunneling_timescale}. Using the explicit expressions for $\gamma(x, \epsilon)$ and $g(x, \epsilon)$,
\begin{widetext}
\begin{equation} \label{eq:REM_tunneling_exponent}
r(\epsilon) = \min_{x \in [x^{**}(\epsilon), 1 - x^{**}(\epsilon)]} \left[ -2x \ln{\frac{\Gamma}{e |\epsilon|}} - x \ln{x} + (1 - x) \ln{(1 - x)} + \epsilon^2 \right] .
\end{equation}
\end{widetext}
It is straightforward to evaluate Eq.~\eqref{eq:REM_tunneling_exponent} numerically.
Fig.~\ref{fig:tunneling_rates} shows the tunneling timescale as a function of $\Gamma$ at representative values of $\epsilon$, alongside the activation timescale.
Tunneling is exponentially faster than thermal activation at energy matching.
One can check that this is true for all $\epsilon$ and $\Gamma$ in the tunneling phase of the REM.
However, in general the tunneling phase may have both a regime in which tunneling is slower than activation (small $\Gamma$) and a regime in which tunneling is faster (large $\Gamma$).
Just as in the single-particle setting, activation timescales depend on the height of energy barriers whereas tunneling timescales depend on a combination of the height and width.
Either can be faster depending on the details of the energy landscape.

\subsection{Numerical results} \label{subsec:numerics}

\begin{figure*}[t]
\begin{center}
\includegraphics[width=1.0\textwidth]{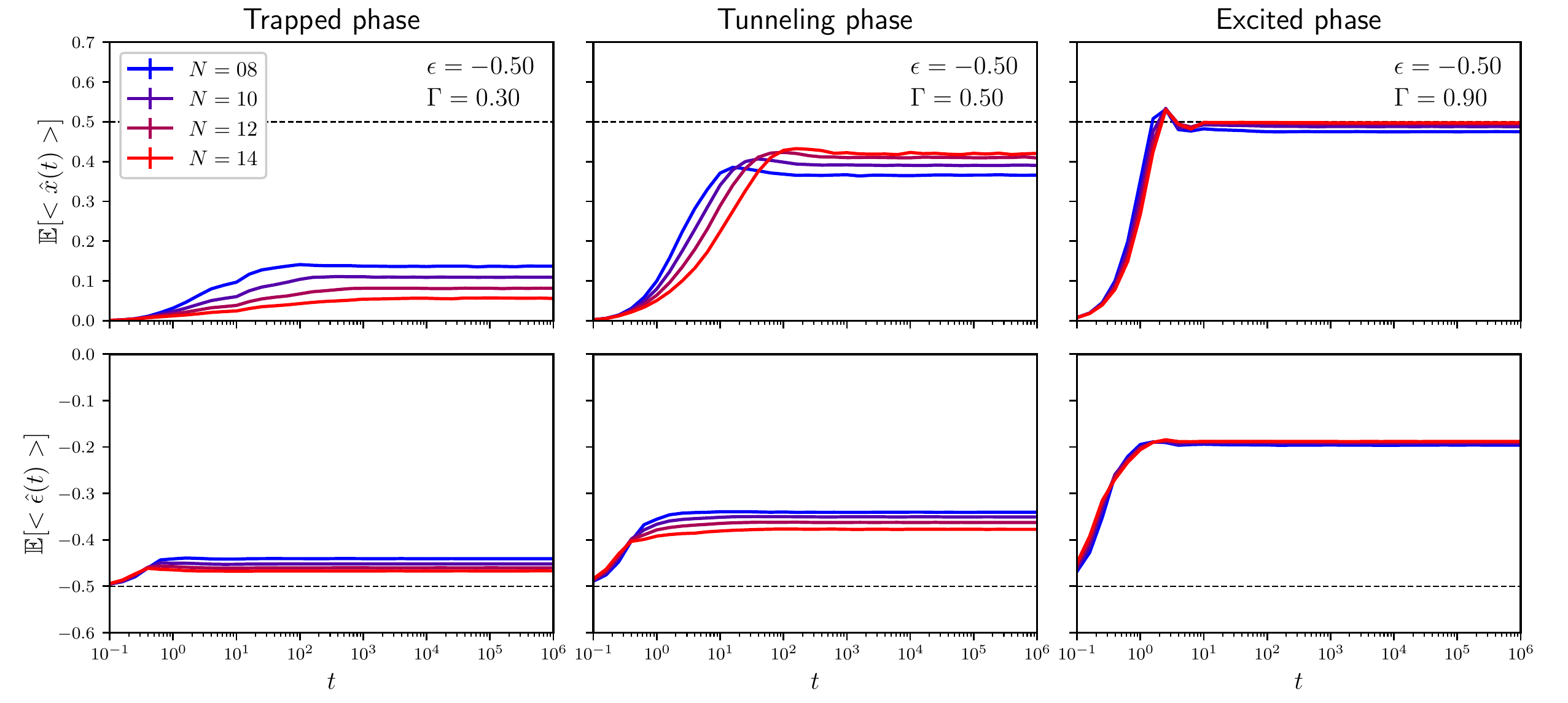}
\caption{Quantum dynamics in the REM, starting from a classical state at energy density $\epsilon$. (Top) Average distance relative to the initial configuration, $\mathbb{E} \big[ \langle \hat{x}(t) \rangle \big] \equiv \mathbb{E} \big[ \braket{\Psi(t) | \hat{x} | \Psi(t)} \big]$. Three representative $\Gamma$ at fixed $\epsilon$ are shown. The dashed line is $x = 1/2$. (Bottom) Average classical energy density, $\mathbb{E} \big[ \langle \hat{\epsilon}(t) \rangle \big] \equiv \mathbb{E} \big[ \braket{\Psi(t) | \hat{\epsilon} | \Psi(t)} \big]$. The same $\Gamma$ and $\epsilon$ as in the top panels are used. The dashed line is the initial energy density $\epsilon$. Statistical errorbars are smaller than the linewidths in all panels. The ``tunneling phase'' label refers to the observed behavior, not the location relative to the perturbatively-obtained phase boundaries in Fig.~\ref{fig:REM_phase_diagram} (see the discussion in Sec.~\ref{subsec:numerics}).}
\label{fig:REM_dynamics_low}
\end{center}
\end{figure*}

As noted above, we made uncontrolled approximations in deriving Eqs.~\eqref{eq:tunneling_condition} and~\eqref{eq:excitation_condition}, namely using the Schrieffer-Wolff transformation and truncating the perturbation series via the FSA.
Thus we use exact diagonalization (ED) studies of the REM to confirm that the dynamical phases in Fig.~\ref{fig:REM_phase_diagram} exist.

Closely related work was performed in Ref.~\cite{Baldwin2016Many}, in the context of eigenstate phases of the quantum REM.
The tunneling transition is also a transition between localized and extended eigenstates in the configuration space, and the authors used multiple measures of localization to detect the transition numerically using ED.
They also compared the location of the transition obtained by ED to the location obtained by evaluating the forward-scattering expression for $\braket{\sigma' | V_{\textrm{eff}} | \sigma}$ numerically.
They found very good agreement, supporting the validity of Schrieffer-Wolff and the FSA. 
Here we supplement these results by studying explicitly \textit{dynamical} properties of the quantum REM.

We construct instances of $H \equiv H_{\textrm{REM}} - \Gamma \sum_i \hat{\sigma}_i^x$, where $H_{\textrm{REM}}$ is diagonal in the $\hat{\sigma}^z$ basis with entries independently distributed according to Eq.~\eqref{eq:REM_energy_distribution}.
We then perform a ``quench'' simulation: find the $\hat{\sigma}^z$ eigenstate $\ket{\sigma}$ with classical energy density closest to a specified $\epsilon$, then compute $\ket{\Psi(t)} \equiv e^{-iHt} \ket{\sigma}$.
We evaluate two observables at time $t$, denoted $\hat{x}$ and $\hat{\epsilon}$, which are the operators corresponding respectively to the distance from $\ket{\sigma}$ and the classical energy density (both are diagonal in the $\hat{\sigma}^z$ basis).
Since we evaluate the full $2^N \times 2^N$ matrix $e^{-iHt}$ via exact diagonalization, we are limited to very small system sizes, namely $N \leq 14$.

Results for $\braket{\Psi(t) | \hat{x} | \Psi(t)}$ and $\braket{\Psi(t) | \hat{\epsilon} | \Psi(t)}$ are shown in Fig.~\ref{fig:REM_dynamics_low}.
Both quantities reach saturated values, denoted $x_{\infty}$ and $\epsilon_{\infty}$.
For all $\Gamma$ and $\epsilon$, $x_{\infty}$ lies between 0 and 1/2 and $\epsilon_{\infty}$ lies between $\epsilon$ and 0.
Since we study finite-size systems, the dependence on the parameters is smooth.
The difference between the three phases is in the flow of $x_{\infty}$ and $\epsilon_{\infty}$ as $N$ increases.
\begin{itemize}
\item Trapped phase: $x_{\infty} \rightarrow 0$ and $\epsilon_{\infty} \rightarrow \epsilon$.
\item Tunneling phase: $x_{\infty} \rightarrow 1/2$ and $\epsilon_{\infty} \rightarrow \epsilon$.
\item Excited phase: $x_{\infty} \rightarrow 1/2$ and $\epsilon_{\infty} \not \rightarrow \epsilon$.
\end{itemize}
In Fig.~\ref{fig:REM_dynamics_low}, the behavior of the system is consistent with the trapped phase at small $\Gamma$ (left panel), the tunneling phase at intermediate $\Gamma$ (middle panel), and the excited phase at large $\Gamma$ (right panel).
Furthermore, it appears that the timescale on which the system approaches $x_{\infty}$ is indeed exponential with $N$ in the tunneling phase.

\begin{figure}[t]
\begin{center}
\includegraphics[width=1.0\columnwidth]{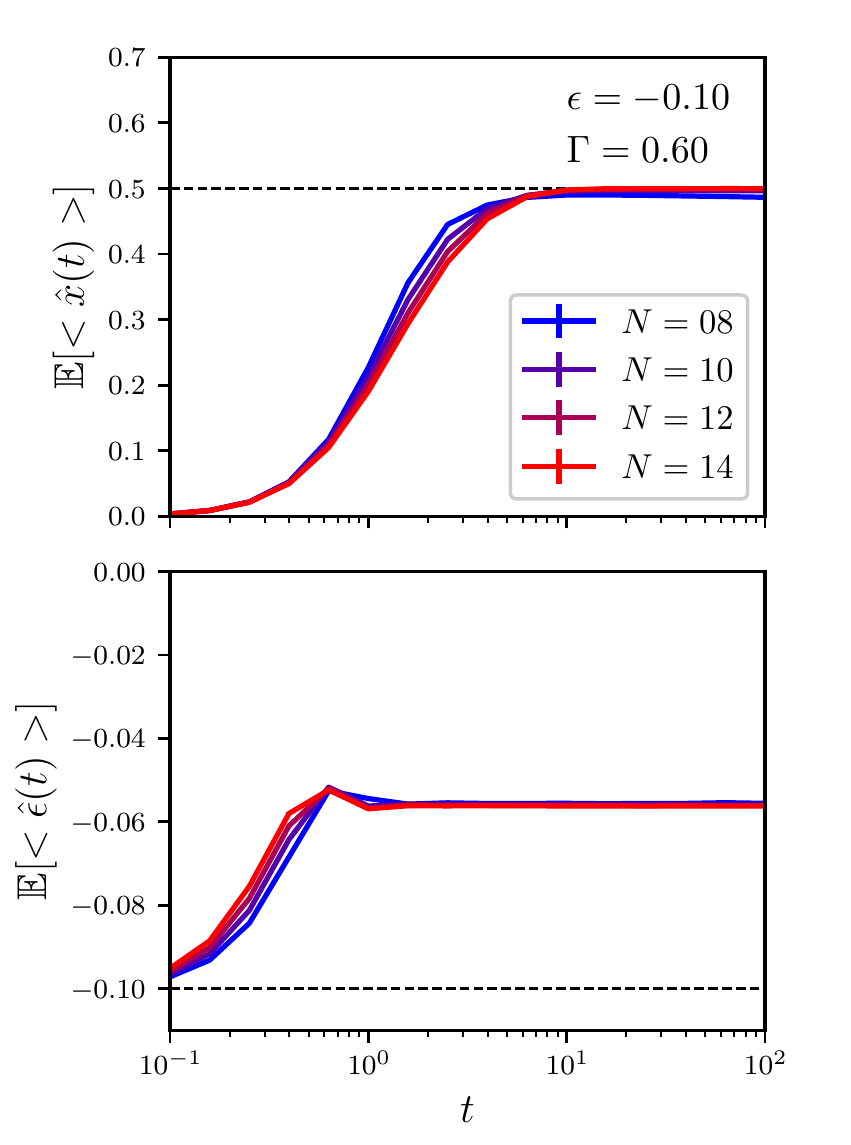}
\caption{Quantum dynamics in the REM starting from a high-energy classical state. (Top) Average distance relative to the initial configuration, $\mathbb{E} \big[ \langle \hat{x}(t) \rangle \big] \equiv \mathbb{E} \big[ \braket{\Psi(t) | \hat{x} | \Psi(t)} \big]$. The dashed line is $x = 1/2$. (Bottom) Average classical energy density, $\mathbb{E} \big[ \langle \hat{\epsilon}(t) \rangle \big] \equiv \mathbb{E} \big[ \braket{\Psi(t) | \hat{\epsilon} | \Psi(t)} \big]$. The dashed line is the initial energy density $\epsilon$. Statistical errorbars are smaller than the linewidths in both panels.}
\label{fig:REM_dynamics_high}
\end{center}
\end{figure}

The middle panel, which shows behavior consistent with the tunneling phase, is at a $(\Gamma, \epsilon)$ point which Fig.~\ref{fig:REM_phase_diagram} predicts is in the trapped phase.
There are two possible explanations for the discrepancy.
First, we cannot rule out that the system becomes trapped at larger system sizes.
Second, since the analysis of Sec.~\ref{subsec:perturbative_analysis} is based on perturbation theory in $\Gamma$, it may only be quantitatively correct at small $\Gamma$.

We are unable to conclusively identify the small-$\Gamma$ excited phase shown in Fig.~\ref{fig:REM_phase_diagram}.
Fig.~\ref{fig:REM_dynamics_high} plots $\braket{\Psi(t) | \hat{x} | \Psi(t)}$ and $\braket{\Psi(t) | \hat{\epsilon} | \Psi(t)}$ for a $(\Gamma, \epsilon)$ point well within that region of the phase diagram.
$x_{\infty} = 1/2$ as expected, signifying that the system is not trapped within a cluster.
However, $\epsilon_{\infty}$ shows a slight downward flow as $N$ increases, and it is not clear whether the timescale on which the system approaches $x_{\infty}$ is scaling exponentially or sub-exponentially.
It is possible that the system is excited out of clusters, but it may instead be in the tunneling phase with very strong finite-size effects.

Regardless, we do find strong signatures of all three dynamical phases at lower $\epsilon$, as shown in Fig.~\ref{fig:REM_dynamics_low}.

\section{Finite-p corrections} \label{sec:finite_p}

\begin{figure}[t]
\begin{center}
\includegraphics[width=1.0\columnwidth]{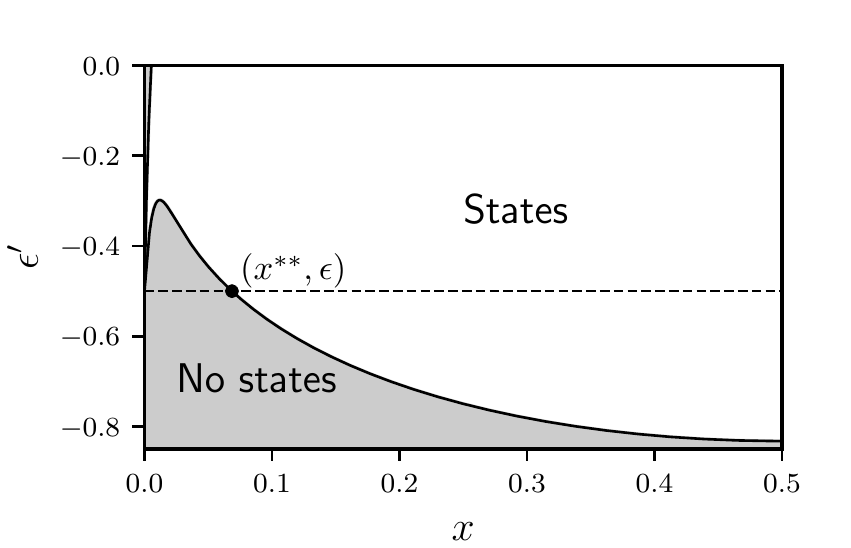}
\caption{Regions of the $(x, \epsilon')$ plane in which the FPP is positive (white) and negative (grey), for $p = 100$ and $\epsilon = -0.5$ (dashed line). Configurations at distance $x$ have energy densities exclusively in the white region. $x^{**}(\epsilon)$ is marked, and $x^*(\epsilon)$ is too close to 0 to be visible on this scale. The curve separating white and grey is $\epsilon'_-(x)$, Eq.~\eqref{eq:energy_barrier_shape} (the corresponding upper root $\epsilon'_+(x)$ is barely visible in the top-left corner).}
\label{fig:spectrum}
\end{center}
\end{figure}

Here we argue that the picture developed above in the limit of vanishing cluster size remains intact for small but non-zero cluster sizes, by considering the leading-in-$1/p$ corrections to the analysis above and showing that they are indeed small.
The additional complexity is that clusters no longer consist of isolated configurations and the $\epsilon(\sigma'')$ that enter into $V_{\textrm{eff}}$ are no longer independent.

\subsection{Geometry of clusters} \label{subsec:cluster_geometry}

First we consider the geometry of clusters at large but finite $p$.
The radius of a cluster is no larger than $x^*(\epsilon)$, where $x^*(\epsilon)$ is the smallest non-zero root of the equation $g(x, \epsilon) = 0$ (see Fig.~\ref{fig:entropy_curves}).
$g(x, \epsilon)$ is given by Eq.~\eqref{eq:microcanonical_fpp_result}, from which we find that
\begin{equation} \label{eq:cluster_radius}
x^*(\epsilon) \sim e^{-O(p)},
\end{equation}
\begin{equation} \label{eq:cluster_separation}
x^{**}(\epsilon) \sim x^{**}(\epsilon)_{p \rightarrow \infty} - e^{-O(p)},
\end{equation}
where $x^{**}(\epsilon)_{p \rightarrow \infty}$ is the REM value of $x^{**}(\epsilon)$.
Thus the distance between clusters is much larger than their radii at large but finite $p$.
The macroscopic energy barriers which separate clusters are also quantified by the FPP.
Setting $g(x, \epsilon' | \epsilon) = 0$ at \textit{fixed} $x$ gives us a bound $\epsilon'_-(x)$: all configurations at distance $x$ have energy densities greater than $\epsilon'_-(x)$.
For $x$ between $x^*(\epsilon)$ and $x^{**}(\epsilon)$,
\begin{widetext}
\begin{equation} \label{eq:energy_barrier_shape}
\epsilon'_-(x) = (1 - 2x)^p \epsilon - \sqrt{\big( 1 - (1 - 2x)^{2p} \big) \big( -x \ln{x} - (1 - x) \ln{(1 - x)} \big)},
\end{equation}
\end{widetext}
which is strictly greater than $\epsilon$.
The shape of $\epsilon'_-(x)$ is plotted in Fig.~\ref{fig:spectrum}.
At large $p$, the peak of the barrier is at
\begin{equation} \label{eq:barrier_peak_position}
x_{\textrm{peak}} \sim \frac{\ln{p}}{4p},
\end{equation}
with a height of
\begin{equation} \label{eq:barrier_peak_height}
\epsilon'_-(x_{\textrm{peak}}) - \epsilon \sim |\epsilon| - \frac{\ln{p}}{2 \sqrt{p}}.
\end{equation}

Given that each cluster has a non-zero radius, and thus that the tunneling paths between $\ket{\sigma}$ and $\ket{\sigma'}$ have portions lying within the clusters, the forward-scattering expansion of $\braket{\sigma' | V_{\textrm{eff}} | \sigma}$ is more complicated than what is given in Eq.~\eqref{eq:forward_scattering_expression} for the REM.
Recall that the REM expression is simply a sum over paths, each term of which takes the same form:
\begin{equation}
\braket{\sigma' | V_{\textrm{eff}} | \sigma} \sim \sum_P \Gamma \prod_{\sigma'' \in P} \frac{\Gamma}{N \big( \epsilon - \epsilon(\sigma'') \big) }. \tag{\ref{eq:forward_scattering_expression}}
\end{equation}
At finite $p$, however, paths acquire different weights and different energy denominators depending on how they pass within clusters (see Appendix~\ref{appendix:schrieffer_wolff}).
We are unable to quantitatively account for these intra-cluster contributions to $\braket{\sigma' | V_{\textrm{eff}} | \sigma}$.
However, we do not expect them to modify the qualitative picture developed for the REM, since the cluster size is much smaller than the separation between clusters.
The majority of each tunneling path lies outside of the clusters and contributes factors of $\Gamma / N \big( \epsilon - \epsilon(\sigma'') \big)$ exactly as in Eq.~\eqref{eq:forward_scattering_expression}.
We expect the intra-cluster portions to give additive contributions to $\gamma(x, \epsilon' | \epsilon)$ which scale as $O \big( x^*(\epsilon) \big)$.
Since $x^*(\epsilon)$ is exponentially small in $p$, these effects are negligible compared to those corrections which we describe below, and we shall continue to use Eq.~\eqref{eq:forward_scattering_expression} for $\braket{\sigma' | V_{\textrm{eff}} | \sigma}$.

We expect the shifts in energy that arise from diagonalizing the intra-cluster effective Hamiltonians ($H_{\textrm{eff}}^{(\alpha)}$ in Fig.~\ref{fig:Schrieffer_Wolff}) to be sub-leading as well.
The number of states within a cluster scales as
\begin{equation} \label{eq:number_cluster_states}
\exp{\left( N \max_{x \in [ 0, x^*(\epsilon) ]} g(x, \epsilon) \right)} \sim \exp{\left( N e^{-O(p)} \right)}.
\end{equation}
Assuming that the hybridization energy among these states scales as the logarithm of Eq.~\eqref{eq:number_cluster_states}, we find that the energy \textit{densities} shift by amounts exponentially small in $p$.
We neglect this effect in what folllows.

\subsection{Intercluster tunneling} \label{subsec:intercluster_tunneling}

The second difference between the finite-$p$ models and the REM is that the factors of $\Gamma / N \big( \epsilon - \epsilon(\sigma'') \big)$ are no longer independent.
We make the approximation
\begin{equation} \label{eq:intermediate_energy_approx}
\epsilon(\sigma'') \approx \mathbb{E} \big[ \epsilon(\sigma'') \big] _{\substack{\epsilon(\sigma) = \epsilon \\ \epsilon(\sigma') = \epsilon'}},
\end{equation}
which accounts for the correlations between $\sigma''$ and the two endpoints but neglects further correlations among the path amplitudes.
The error in making this approximation is only $O(1/p)$, as we now show.

Consider those tunneling paths along which the energy densities are a given function $\epsilon(y)$ ($0 < y < x$).
Let the number of such paths be denoted $\mathcal{N} \big( \epsilon(y) \big)$.
We can then write Eq.~\eqref{eq:forward_scattering_expression} as a path integral over $\epsilon(y)$:
\begin{equation} \label{eq:forward_scattering_formal_rewrite}
\big| \braket{\sigma' | V_{\textrm{eff}} | \sigma} \big| \sim \int \mathcal{D}\epsilon(y) \, \mathcal{N} \big( \epsilon(y) \big) \, e^{N \int_0^x \textrm{d}y \, \ln{\frac{\Gamma}{N|\epsilon - \epsilon(y)|}}}.
\end{equation}
We enforce the conditioning on $\epsilon(\sigma)$ and $\epsilon(\sigma')$ through boundary conditions: $\epsilon(y = 0) = \epsilon$ and $\epsilon(y = x) = \epsilon'$.
Eq.~\eqref{eq:forward_scattering_formal_rewrite} is formally exact, since now $\mathcal{N} \big( \epsilon(y) \big)$ is the random variable which depends on the disorder realization.
Importantly, we integrate only over $\epsilon(y)$ that are nowhere equal to $\epsilon$, since the tunneling paths lie outside of the clusters.
We next make another ``annealed'' approximation:
\begin{equation} \label{eq:path_number_annealed}
\mathcal{N} \big( \epsilon(y) \big) \rightarrow \mathbb{E} \big[ \mathcal{N} \big( \epsilon(y) \big) \big] = \big( Nx \big) ! \, P \big( \epsilon(y) \big),
\end{equation}
where $P \big( \epsilon(y) \big)$ is the \textit{probability} of a given path having energy densities $\epsilon(y)$.
Calculating $P \big( \epsilon(y) \big)$ exactly is intractable, yet we can infer the scaling with both $N$ and $p$ from the covariance matrix of the classical energies (Eq.~\eqref{eq:p_spin_cov_matrix}):
\begin{equation} \label{eq:p_spin_alt_cov_matrix}
\mathbb{E} \big[ \epsilon(\sigma_1) \epsilon(\sigma_2) \big] = \frac{1}{2N} \left( 1 - 2x(\sigma_1, \sigma_2) \right) ^p.
\end{equation}
At large $p$, the right-hand side is independent of $p$ on a length scale $x(\sigma_1, \sigma_2) \sim O \big( 1/p \big)$.
Thus $1/p$ can be identified as the ``correlation length'' for the classical energies.
If $\epsilon(y)$ deviates from its mean throughout a distance $x$, then since the number of correlation lengths involved is $px$, this roughly corresponds to $px$ independent fluctuations, each of which is exponentially rare in $N$.
Thus
\begin{equation} \label{eq:path_probability_scaling}
P \big( \epsilon(y) \big) \sim e^{-N p \, c ( \epsilon(y) )},
\end{equation}
where $c \big( \epsilon(y) \big)$ is independent of both $N$ and $p$, and $c \big( \mathbb{E} \big[ \epsilon(y) \big] \big) = 0$.
Thus Eq.~\eqref{eq:forward_scattering_formal_rewrite} becomes
\begin{widetext}
\begin{equation} \label{eq:forward_scattering_scaling}
\big| \braket{\sigma' | V_{\textrm{eff}} | \sigma} \big| \sim \int \mathcal{D}\epsilon(y) \, e^{N \left( x \ln{\frac{x \Gamma}{e}} - \int_0^x \textrm{d}y \, \ln{|\epsilon - \epsilon(y)|} - p \, c ( \epsilon(y) ) \right)}.
\end{equation}
At $p \rightarrow \infty$, the saddle point of the path integral is at $\epsilon^0(y) \equiv \mathbb{E} \big[ \epsilon(y) \big]$ (see Sec.~\ref{subsec:perturbative_analysis}).
To compute the correction from large but finite $p$, write $\epsilon(y) = \epsilon^0(y) + \delta \epsilon(y)$ and expand the exponent:
\begin{equation} \label{eq:path_saddle_point_correction}
- \int_0^x \textrm{d}y \, \ln{|\epsilon - \epsilon(y)|} - p \, c ( \epsilon(y) ) \sim - \int_0^x \textrm{d}y \, \ln{|\epsilon - \epsilon^0(y)|} \, + \, \int_0^x \textrm{d}y \, \frac{\delta \epsilon(y)}{\epsilon - \epsilon(y)} \, - \, p \, \frac{\partial^2 c \big( \epsilon^0(y) \big)}{\partial \epsilon(y) \partial \epsilon(z)} \delta \epsilon(y) \delta \epsilon(z).
\end{equation}
There are no first derivatives of $c \big( \epsilon(y) \big)$ because $\epsilon^0(y)$ is the location of its minimum.
Competition between the second and third terms in Eq.~\eqref{eq:path_saddle_point_correction} determines the location of the saddle point.
They are comparable for $\delta \epsilon(y) \sim O \big( 1/p \big)$, which changes the value of the exponent at the saddle point by $O \big( 1/p \big)$.
This is the error made to $\braket{\sigma' | V_{\textrm{eff}} | \sigma}$, or rather $\gamma(x, \epsilon' | \epsilon)$, by approximating $\epsilon(y) \approx \mathbb{E} \big[ \epsilon(y) \big]$, i.e., $\delta \epsilon(y) \approx 0$.
It is indeed small at large $p$.

Next consider how $\mathbb{E} \big[ \epsilon(y) \big]$ is modified by correlations with $\epsilon(\sigma)$ and $\epsilon(\sigma')$.
A straightforward calculation given in Appendix~\ref{appendix:energy_correlations} shows that
\begin{equation} \label{eq:intermediate_average_energy}
\mathbb{E} \big[ \epsilon(y) \big] \bigg|_{\substack{\epsilon(\sigma) = \epsilon \\ \epsilon(\sigma') = \epsilon'}} = \frac{\mu(y) - \mu(x) \mu(x - y)}{1 - \mu(x)^2} \epsilon \, + \, \frac{\mu(x - y) - \mu(x) \mu(y)}{1 - \mu(x)^2} \epsilon' ,
\end{equation}
where $\mu(z) \equiv (1 - 2z)^p$.
Then $\big| \braket{\sigma' | V_{\textrm{eff}} | \sigma} \big| \sim e^{-N \gamma(x, \epsilon' | \epsilon)}$ with
\begin{equation} \label{eq:finite_p_effective_coupling_exponent}
\gamma(x, \epsilon' | \epsilon) = -x \ln{\frac{x \Gamma}{e |\epsilon|}} + \int_0^x \textrm{d}y \, \ln{\left| 1 - \frac{\mu(y) - \mu(x) \mu(x - y)}{1 - \mu(x)^2} - \frac{\mu(x - y) - \mu(x) \mu(y)}{1 - \mu(x)^2} \frac{\epsilon'}{\epsilon} \right| }.
\end{equation}
In the large-$p$ limit, assuming $x \sim O(1)$~\footnote{To obtain Eq.~\eqref{eq:effective_coupling_large_p_expansion} from Eq.~\eqref{eq:finite_p_effective_coupling_exponent}, substitute $y = w/p$ and take $\mu(w/p) \sim e^{-2w}$.},
\begin{equation} \label{eq:effective_coupling_large_p_expansion}
\gamma(x, \epsilon' | \epsilon) \sim -x \ln{\frac{x \Gamma}{e |\epsilon|}} - \frac{K}{p} ,
\end{equation}
where
\begin{equation} \label{eq:effective_coupling_expansion_integral}
K \equiv \frac{\pi^2}{12} - \int_0^{\infty} \textrm{d}y \, \ln{ \left( 1 - \frac{\epsilon'}{\epsilon} e^{-2y} \right) }.
\end{equation}
The correction relative to Eqs.~\eqref{eq:REM_effective_coupling_exponent} and~\eqref{eq:outside_amplitude_evaluation} for the REM is indeed $O \big( 1/p \big)$.

\subsection{Phase boundaries} \label{subsec:phase_boundaries}

The same arguments as in Sec.~\ref{subsec:perturbative_analysis} give Eq.~\eqref{eq:tunneling_condition} as a necessary condition for tunneling between clusters to occur:
\begin{equation}
\max_{x \in [x^{**}(\epsilon), 1 - x^{**}(\epsilon)]} \big[ g(x, \epsilon) - \gamma(x, \epsilon) \big] > 0, \tag{\ref{eq:tunneling_condition}}
\end{equation}
only now with Eq.~\eqref{eq:microcanonical_fpp_result} for $g(x, \epsilon)$ and Eq.~\eqref{eq:effective_coupling_large_p_expansion} for $\gamma(x, \epsilon)$.
Since $x^{**}(\epsilon) \sim O(1)$ with respect to $p$, $g(x, \epsilon)$ differs from the REM expression (Eq.~\eqref{eq:REM_microcanonical_fpp}) by an amount exponentially small in $p$ and $\gamma(x, \epsilon)$ differs from Eq.~\eqref{eq:REM_effective_coupling_exponent} by $O \big( 1/p \big)$.
Thus the same Eq.~\eqref{eq:REM_tunneling_condition} determines the tunneling transition in the finite-$p$ models to within $O \big( 1/p \big)$.

The excitation transition in the finite-$p$ models is more subtle.
Within the approximations that we have made, Eq.~\eqref{eq:finite_p_effective_coupling_exponent} holds for all $x$.
However, Eq.~\eqref{eq:effective_coupling_large_p_expansion} only holds for $x \sim O(1)$.
At $x \ll 1/p$, we instead have
\begin{equation} \label{eq:finite_p_effective_coupling_small_x}
\gamma(x, \epsilon' | \epsilon) \sim -x \ln{\frac{e \Gamma}{2p^2 |\epsilon|}} + \left( x + \frac{\epsilon' - \epsilon}{2p^2 |\epsilon| x} \right) \ln{\left( x + \frac{\epsilon' - \epsilon}{2p^2 |\epsilon| x} \right)} - \frac{\epsilon' - \epsilon}{2p^2 |\epsilon| x} \ln{\frac{\epsilon' - \epsilon}{2p^2 |\epsilon| x}},
\end{equation}
\begin{equation} \label{eq:finite_p_excitation_quantity_small_x}
g(x, \epsilon' | \epsilon) - 2 \gamma(x, \epsilon' | \epsilon) \sim x \ln{\frac{e^3 \Gamma^2}{4p^4 \epsilon^2 x}} - \frac{\big( \epsilon' - \epsilon - 2p|\epsilon|x \big) ^2}{4px} - 2 \left( x + \frac{\epsilon' - \epsilon}{2p^2 |\epsilon| x} \right) \ln{\left( x + \frac{\epsilon' - \epsilon}{2p^2 |\epsilon| x} \right)} + \frac{\epsilon' - \epsilon}{p^2 |\epsilon| x} \ln{\frac{\epsilon' - \epsilon}{2p^2 |\epsilon| x}}.
\end{equation}
\end{widetext}
The maximum over $\epsilon'$ is at $\epsilon' - \epsilon = 2p |\epsilon| x + O \big( 1/p \big)$, and
\begin{equation} \label{eq:finite_p_excitation_energy_max}
\max_{\epsilon'} \big[ g(x, \epsilon' | \epsilon) - 2 \gamma(x, \epsilon' | \epsilon) \big] = x \ln{\frac{e^3 \Gamma^2}{4p^2 \epsilon^2 x}} + O \left( \frac{1}{p^3} \right) .
\end{equation}
The maximum over $x$ is then at
\begin{equation} \label{eq:finite_p_wavefunction_tail}
x_m = \frac{e^2 \Gamma^2}{4p^2 \epsilon^2},
\end{equation}
and importantly, the value of the maximum is positive.
Eq.~\eqref{eq:excitation_condition} is never satisfied, and the system is excited to higher classical energy densities.
However, this does not imply that the system escapes from its starting cluster: Eq.~\eqref{eq:finite_p_wavefunction_tail} indicates that the system moves through a distance $O \big( 1/p^2 \big)$, and thus the classical energy changes by $O \big( 1/p \big)$, which is much too small to surmount the energy barrier at distance $O \big( 1/p \big)$ and height $O(1)$ (Fig.~\ref{fig:spectrum} and Eq.~\eqref{eq:barrier_peak_position}).
To check if the system is excited out of clusters, one should modify Eq.~\eqref{eq:excitation_condition} to maximize over $x \gtrsim O \big( 1/p \big)$.
Then the finite-$p$ corrections to $\gamma(x, \epsilon' | \epsilon)$ are subleading, and Eq.~\eqref{eq:excitation_condition} determines the excitation transition to within $O \big( 1/p \big)$.

The thermodynamic transition into a quantum paramagnetic phase, constituting another excitation transition, is also present at finite $p$~\cite{Dobrosavljevic1990Expansion,Nieuwenhuizen19988}.
The phase boundary terminates at energy density $-O \big( 1/\sqrt{p} \big)$~\cite{Dobrosavljevic1990Expansion}.
On the other hand, by setting $\min_x \big[ g(x, \epsilon) \big] = 0$, one finds that
\begin{equation} \label{eq:epsilon_d}
\epsilon_d \sim - \sqrt{\frac{\ln{p}}{p}}.
\end{equation}
Since the energy landscape of the $p$-spin model is organized into clusters only below $\epsilon_d$, the large-$\Gamma$ excitation transition is present at all $\epsilon$ for which energy matching is non-trivial.

\section{Conclusion} \label{sec:conclusion}

We have studied the performance of quantum dynamics in energy matching problems, using perturbation theory and exact diagonalization applied to the $p$-spin models at large $p$.
The goal of energy matching is to find states at a target energy given one state at the same energy.
In general, finding low energy states is difficult when the energy landscape is rugged, i.e., contains many local minima separated by large barriers.
Ideally, starting at a low energy state serves as a ``hint'' for finding others, but the same ruggedness makes energy matching difficult: 
in order to find sufficiently distinct target states, any classical algorithm that flips only a few spins per step must return to and explore high energy states at intermediate steps.
This raises the possibility for quantum dynamics to provide a speed-up over classical algorithms by tunneling through those energy barriers.

In many regimes of transverse field strength and target energy, quantum dynamics \textit{cannot} succeed in energy matching even if allowed arbitrarily long runtime.
If the applied field is too weak, the system never tunnels out of its starting cluster of states.
If the applied field is too strong, the system is excited to higher (classical) energy densities and thus never locates target states.
Only at intermediate fields does the system successfully tunnel between clusters while roughly conserving the classical energy density.
These possibilities constitute distinct dynamical phases, demarcated by sharp phase boundaries in the thermodynamic limit (Fig.~\ref{fig:REM_phase_diagram}).
The combination of three features of the energy landscape underlies the physics: the macroscopic height of energy barriers, the macroscopic width of those barriers, and the exponential number of clusters in which to tunnel (referred to in the spin glass community as the ``complexity'' of the model).
These features are typical in many of the well-studied classes of optimization problems~\cite{Biroli2000Variational,Mezard2002Random,Mezard2003Two,Zdeborova2007Phase}, thus we expect the results of this paper to apply more generally.

A remarkable feature of the dynamical phase diagram for the $p$-spin model is that there is no tunneling phase at energies close to the ground state.
The barriers between such low energy states are too large and wide, and the complexity of clusters is too small.
As a result, it is not possible to tunnel between ground or near-ground states of the $p$-spin model for any applied transverse field.
However, ground state tunneling may be possible in problems with larger ground state complexity, e.g. the satisfiability problem~\cite{Mezard2002Analytic,Kirkpatrick1994Critical,Bapst2013Quantum}, and we leave this as an interesting open question.

If the system is in its tunneling phase, the timescale for intercluster tunneling is exponential in system size.
Thus quantum dynamics never solves the energy matching problem in polynomial time.
We do find that tunneling is exponentially faster than certain simple classical algorithms, such as Metropolis Monte Carlo simulation, throughout the tunneling phase.
One drawback is that the rates we find are exponentially slower than those which could be achieved by a universal quantum computer implementing a variation of Grover's algorithm~\cite{Nielsen2000Quantum}.
Indeed, in the REM ($p \rightarrow \infty$ limit), there are strong arguments that \textit{no} algorithm can outperform Grover's since the energy landscape is unstructured~\cite{Farhi2008How}.
However, Grover's algorithm requires both exponential tuning precision and detailed knowledge of the adiabatic level structure.
The advantage of the algorithm presented here is that it requires neither.
Effective Hamiltonians of the form in Eq.~\eqref{eq:Hamiltonian_form} are realizable in near-term quantum annealing devices~\cite{Johnson2011Quantum,Boixo2014Evidence}, and the energy matching protocol proposed here is straightforward to implement and test on them.

\section{Acknowledgements}

We would like to thank for A. Chandran, D. Huse, F. Krzakala, A. Pal, F. Ricci-Tersenghi, O. Sattath, A. Scardicchio, and V. Smelyanskiy for useful discussions.
We would also like to thank the Simons Center for Geometry and Physics and the Kavli Institute for Theoretical Physics for hospitality during which part of this work was completed.
CLB acknowledges the support of the NSF through a Graduate Research Fellowship, Grant No. DGE-1256082.
CRL acknowledges support from the Sloan Foundation through a Sloan Research Fellowship and the NSF through Grant No. PHY-1656234.

\bibliography{PSD_Biblio_v3}

\begin{thebibliography}{48}%
\makeatletter
\providecommand \@ifxundefined [1]{%
 \@ifx{#1\undefined}
}%
\providecommand \@ifnum [1]{%
 \ifnum #1\expandafter \@firstoftwo
 \else \expandafter \@secondoftwo
 \fi
}%
\providecommand \@ifx [1]{%
 \ifx #1\expandafter \@firstoftwo
 \else \expandafter \@secondoftwo
 \fi
}%
\providecommand \natexlab [1]{#1}%
\providecommand \enquote  [1]{``#1''}%
\providecommand \bibnamefont  [1]{#1}%
\providecommand \bibfnamefont [1]{#1}%
\providecommand \citenamefont [1]{#1}%
\providecommand \href@noop [0]{\@secondoftwo}%
\providecommand \href [0]{\begingroup \@sanitize@url \@href}%
\providecommand \@href[1]{\@@startlink{#1}\@@href}%
\providecommand \@@href[1]{\endgroup#1\@@endlink}%
\providecommand \@sanitize@url [0]{\catcode `\\12\catcode `\$12\catcode
  `\&12\catcode `\#12\catcode `\^12\catcode `\_12\catcode `\%12\relax}%
\providecommand \@@startlink[1]{}%
\providecommand \@@endlink[0]{}%
\providecommand \url  [0]{\begingroup\@sanitize@url \@url }%
\providecommand \@url [1]{\endgroup\@href {#1}{\urlprefix }}%
\providecommand \urlprefix  [0]{URL }%
\providecommand \Eprint [0]{\href }%
\providecommand \doibase [0]{http://dx.doi.org/}%
\providecommand \selectlanguage [0]{\@gobble}%
\providecommand \bibinfo  [0]{\@secondoftwo}%
\providecommand \bibfield  [0]{\@secondoftwo}%
\providecommand \translation [1]{[#1]}%
\providecommand \BibitemOpen [0]{}%
\providecommand \bibitemStop [0]{}%
\providecommand \bibitemNoStop [0]{.\EOS\space}%
\providecommand \EOS [0]{\spacefactor3000\relax}%
\providecommand \BibitemShut  [1]{\csname bibitem#1\endcsname}%
\let\auto@bib@innerbib\@empty
\bibitem [{\citenamefont {Cook}(2012)}]{Cook2012Pursuit}%
  \BibitemOpen
  \bibfield  {author} {\bibinfo {author} {\bibfnamefont {W.~J.}\ \bibnamefont
  {Cook}},\ }\href@noop {} {\emph {\bibinfo {title} {In Pursuit of the
  Traveling Salesman}}}\ (\bibinfo  {publisher} {Princeton University Press},\
  \bibinfo {year} {2012})\BibitemShut {NoStop}%
\bibitem [{\citenamefont {Mezard}\ \emph {et~al.}(2002)\citenamefont {Mezard},
  \citenamefont {Parisi},\ and\ \citenamefont {Zecchina}}]{Mezard2002Analytic}%
  \BibitemOpen
  \bibfield  {author} {\bibinfo {author} {\bibfnamefont {M.}~\bibnamefont
  {Mezard}}, \bibinfo {author} {\bibfnamefont {G.}~\bibnamefont {Parisi}}, \
  and\ \bibinfo {author} {\bibfnamefont {R.}~\bibnamefont {Zecchina}},\
  }\href@noop {} {\bibfield  {journal} {\bibinfo  {journal} {Science}\ }\textbf
  {\bibinfo {volume} {297}},\ \bibinfo {pages} {812} (\bibinfo {year}
  {2002})}\BibitemShut {NoStop}%
\bibitem [{\citenamefont {Mulet}\ \emph {et~al.}(2002)\citenamefont {Mulet},
  \citenamefont {Pagnani}, \citenamefont {Weigt},\ and\ \citenamefont
  {Zecchina}}]{Mulet2002Coloring}%
  \BibitemOpen
  \bibfield  {author} {\bibinfo {author} {\bibfnamefont {R.}~\bibnamefont
  {Mulet}}, \bibinfo {author} {\bibfnamefont {A.}~\bibnamefont {Pagnani}},
  \bibinfo {author} {\bibfnamefont {M.}~\bibnamefont {Weigt}}, \ and\ \bibinfo
  {author} {\bibfnamefont {R.}~\bibnamefont {Zecchina}},\ }\href@noop {}
  {\bibfield  {journal} {\bibinfo  {journal} {Phys. Rev. Lett.}\ }\textbf
  {\bibinfo {volume} {89}},\ \bibinfo {pages} {268701} (\bibinfo {year}
  {2002})}\BibitemShut {NoStop}%
\bibitem [{\citenamefont {Raymond}\ \emph {et~al.}(2007)\citenamefont
  {Raymond}, \citenamefont {Sportiello},\ and\ \citenamefont
  {Zdeborov\'a}}]{Raymond2007Phase}%
  \BibitemOpen
  \bibfield  {author} {\bibinfo {author} {\bibfnamefont {J.}~\bibnamefont
  {Raymond}}, \bibinfo {author} {\bibfnamefont {A.}~\bibnamefont {Sportiello}},
  \ and\ \bibinfo {author} {\bibfnamefont {L.}~\bibnamefont {Zdeborov\'a}},\
  }\href@noop {} {\bibfield  {journal} {\bibinfo  {journal} {Phys. Rev. E}\
  }\textbf {\bibinfo {volume} {76}},\ \bibinfo {pages} {011101} (\bibinfo
  {year} {2007})}\BibitemShut {NoStop}%
\bibitem [{\citenamefont {Krzakala}\ and\ \citenamefont
  {Zdeborov\'a}(2009)}]{Krzakala2009Hiding}%
  \BibitemOpen
  \bibfield  {author} {\bibinfo {author} {\bibfnamefont {F.}~\bibnamefont
  {Krzakala}}\ and\ \bibinfo {author} {\bibfnamefont {L.}~\bibnamefont
  {Zdeborov\'a}},\ }\href@noop {} {\bibfield  {journal} {\bibinfo  {journal}
  {Phys. Rev. Lett.}\ }\textbf {\bibinfo {volume} {102}},\ \bibinfo {pages}
  {238701} (\bibinfo {year} {2009})}\BibitemShut {NoStop}%
\bibitem [{\citenamefont {Baldassi}\ \emph {et~al.}(2016)\citenamefont
  {Baldassi}, \citenamefont {Borgs}, \citenamefont {Chayes}, \citenamefont
  {Ingrosso}, \citenamefont {Lucibello}, \citenamefont {Saglietti},\ and\
  \citenamefont {Zecchina}}]{Baldassi2016Unreasonable}%
  \BibitemOpen
  \bibfield  {author} {\bibinfo {author} {\bibfnamefont {C.}~\bibnamefont
  {Baldassi}}, \bibinfo {author} {\bibfnamefont {C.}~\bibnamefont {Borgs}},
  \bibinfo {author} {\bibfnamefont {J.~T.}\ \bibnamefont {Chayes}}, \bibinfo
  {author} {\bibfnamefont {A.}~\bibnamefont {Ingrosso}}, \bibinfo {author}
  {\bibfnamefont {C.}~\bibnamefont {Lucibello}}, \bibinfo {author}
  {\bibfnamefont {L.}~\bibnamefont {Saglietti}}, \ and\ \bibinfo {author}
  {\bibfnamefont {R.}~\bibnamefont {Zecchina}},\ }\href@noop {} {\bibfield
  {journal} {\bibinfo  {journal} {Proceedings of the National Academy of
  Sciences}\ }\textbf {\bibinfo {volume} {113}},\ \bibinfo {pages} {E7655}
  (\bibinfo {year} {2016})}\BibitemShut {NoStop}%
\bibitem [{\citenamefont {Farhi}\ \emph {et~al.}(2001)\citenamefont {Farhi},
  \citenamefont {Goldstone}, \citenamefont {Gutmann}, \citenamefont {Lapan},
  \citenamefont {Lundgren},\ and\ \citenamefont {Preda}}]{Farhi2001Quantum}%
  \BibitemOpen
  \bibfield  {author} {\bibinfo {author} {\bibfnamefont {E.}~\bibnamefont
  {Farhi}}, \bibinfo {author} {\bibfnamefont {J.}~\bibnamefont {Goldstone}},
  \bibinfo {author} {\bibfnamefont {S.}~\bibnamefont {Gutmann}}, \bibinfo
  {author} {\bibfnamefont {J.}~\bibnamefont {Lapan}}, \bibinfo {author}
  {\bibfnamefont {A.}~\bibnamefont {Lundgren}}, \ and\ \bibinfo {author}
  {\bibfnamefont {D.}~\bibnamefont {Preda}},\ }\href@noop {} {\bibfield
  {journal} {\bibinfo  {journal} {Science}\ }\textbf {\bibinfo {volume}
  {292}},\ \bibinfo {pages} {472} (\bibinfo {year} {2001})}\BibitemShut
  {NoStop}%
\bibitem [{\citenamefont {Knysh}\ and\ \citenamefont
  {Smelyanskiy}(2008)}]{Knysh2008Statistical}%
  \BibitemOpen
  \bibfield  {author} {\bibinfo {author} {\bibfnamefont {S.}~\bibnamefont
  {Knysh}}\ and\ \bibinfo {author} {\bibfnamefont {V.~N.}\ \bibnamefont
  {Smelyanskiy}},\ }\href@noop {} {\bibfield  {journal} {\bibinfo  {journal}
  {Phys. Rev. E}\ }\textbf {\bibinfo {volume} {78}},\ \bibinfo {pages} {061128}
  (\bibinfo {year} {2008})}\BibitemShut {NoStop}%
\bibitem [{\citenamefont {Young}\ \emph {et~al.}(2010)\citenamefont {Young},
  \citenamefont {Knysh},\ and\ \citenamefont {Smelyanskiy}}]{Young2010First}%
  \BibitemOpen
  \bibfield  {author} {\bibinfo {author} {\bibfnamefont {A.~P.}\ \bibnamefont
  {Young}}, \bibinfo {author} {\bibfnamefont {S.}~\bibnamefont {Knysh}}, \ and\
  \bibinfo {author} {\bibfnamefont {V.~N.}\ \bibnamefont {Smelyanskiy}},\
  }\href@noop {} {\bibfield  {journal} {\bibinfo  {journal} {Phys. Rev. Lett.}\
  }\textbf {\bibinfo {volume} {104}},\ \bibinfo {pages} {020502} (\bibinfo
  {year} {2010})}\BibitemShut {NoStop}%
\bibitem [{\citenamefont {Farhi}\ \emph {et~al.}(2012)\citenamefont {Farhi},
  \citenamefont {Gosset}, \citenamefont {Hen}, \citenamefont {Sandvik},
  \citenamefont {Shor}, \citenamefont {Young},\ and\ \citenamefont
  {Zamponi}}]{Farhi2012Performance}%
  \BibitemOpen
  \bibfield  {author} {\bibinfo {author} {\bibfnamefont {E.}~\bibnamefont
  {Farhi}}, \bibinfo {author} {\bibfnamefont {D.}~\bibnamefont {Gosset}},
  \bibinfo {author} {\bibfnamefont {I.}~\bibnamefont {Hen}}, \bibinfo {author}
  {\bibfnamefont {A.~W.}\ \bibnamefont {Sandvik}}, \bibinfo {author}
  {\bibfnamefont {P.}~\bibnamefont {Shor}}, \bibinfo {author} {\bibfnamefont
  {A.~P.}\ \bibnamefont {Young}}, \ and\ \bibinfo {author} {\bibfnamefont
  {F.}~\bibnamefont {Zamponi}},\ }\href@noop {} {\bibfield  {journal} {\bibinfo
   {journal} {Phys. Rev. A}\ }\textbf {\bibinfo {volume} {86}},\ \bibinfo
  {pages} {052334} (\bibinfo {year} {2012})}\BibitemShut {NoStop}%
\bibitem [{\citenamefont {Isakov}\ \emph {et~al.}(2016)\citenamefont {Isakov},
  \citenamefont {Mazzola}, \citenamefont {Smelyanskiy}, \citenamefont {Jiang},
  \citenamefont {Boixo}, \citenamefont {Neven},\ and\ \citenamefont
  {Troyer}}]{Isakov2016Understanding}%
  \BibitemOpen
  \bibfield  {author} {\bibinfo {author} {\bibfnamefont {S.~V.}\ \bibnamefont
  {Isakov}}, \bibinfo {author} {\bibfnamefont {G.}~\bibnamefont {Mazzola}},
  \bibinfo {author} {\bibfnamefont {V.~N.}\ \bibnamefont {Smelyanskiy}},
  \bibinfo {author} {\bibfnamefont {Z.}~\bibnamefont {Jiang}}, \bibinfo
  {author} {\bibfnamefont {S.}~\bibnamefont {Boixo}}, \bibinfo {author}
  {\bibfnamefont {H.}~\bibnamefont {Neven}}, \ and\ \bibinfo {author}
  {\bibfnamefont {M.}~\bibnamefont {Troyer}},\ }\href@noop {} {\bibfield
  {journal} {\bibinfo  {journal} {Phys. Rev. Lett.}\ }\textbf {\bibinfo
  {volume} {117}},\ \bibinfo {pages} {180402} (\bibinfo {year}
  {2016})}\BibitemShut {NoStop}%
\bibitem [{\citenamefont {Altshuler}\ \emph {et~al.}(2010)\citenamefont
  {Altshuler}, \citenamefont {Krovi},\ and\ \citenamefont
  {Roland}}]{Altshuler2010Anderson}%
  \BibitemOpen
  \bibfield  {author} {\bibinfo {author} {\bibfnamefont {B.}~\bibnamefont
  {Altshuler}}, \bibinfo {author} {\bibfnamefont {H.}~\bibnamefont {Krovi}}, \
  and\ \bibinfo {author} {\bibfnamefont {J.}~\bibnamefont {Roland}},\
  }\href@noop {} {\bibfield  {journal} {\bibinfo  {journal} {Proceedings of the
  National Academy of Sciences}\ }\textbf {\bibinfo {volume} {107}},\ \bibinfo
  {pages} {12446} (\bibinfo {year} {2010})}\BibitemShut {NoStop}%
\bibitem [{\citenamefont {Baldwin}\ \emph {et~al.}(2016)\citenamefont
  {Baldwin}, \citenamefont {Laumann}, \citenamefont {Pal},\ and\ \citenamefont
  {Scardicchio}}]{Baldwin2016Many}%
  \BibitemOpen
  \bibfield  {author} {\bibinfo {author} {\bibfnamefont {C.~L.}\ \bibnamefont
  {Baldwin}}, \bibinfo {author} {\bibfnamefont {C.~R.}\ \bibnamefont
  {Laumann}}, \bibinfo {author} {\bibfnamefont {A.}~\bibnamefont {Pal}}, \ and\
  \bibinfo {author} {\bibfnamefont {A.}~\bibnamefont {Scardicchio}},\
  }\href@noop {} {\bibfield  {journal} {\bibinfo  {journal} {Phys. Rev. B}\
  }\textbf {\bibinfo {volume} {93}},\ \bibinfo {pages} {024202} (\bibinfo
  {year} {2016})}\BibitemShut {NoStop}%
\bibitem [{\citenamefont {Baldwin}\ \emph {et~al.}(2017)\citenamefont
  {Baldwin}, \citenamefont {Laumann}, \citenamefont {Pal},\ and\ \citenamefont
  {Scardicchio}}]{Baldwin2017Clustering}%
  \BibitemOpen
  \bibfield  {author} {\bibinfo {author} {\bibfnamefont {C.~L.}\ \bibnamefont
  {Baldwin}}, \bibinfo {author} {\bibfnamefont {C.~R.}\ \bibnamefont
  {Laumann}}, \bibinfo {author} {\bibfnamefont {A.}~\bibnamefont {Pal}}, \ and\
  \bibinfo {author} {\bibfnamefont {A.}~\bibnamefont {Scardicchio}},\
  }\href@noop {} {\bibfield  {journal} {\bibinfo  {journal} {Phys. Rev. Lett.}\
  }\textbf {\bibinfo {volume} {118}},\ \bibinfo {pages} {127201} (\bibinfo
  {year} {2017})}\BibitemShut {NoStop}%
\bibitem [{Note1()}]{Note1}%
  \BibitemOpen
  \bibinfo {note} {Dynamical fluctuations in the classical energy density are
  $O(1/N)$ in the REM, and are $O(1/p)$ in the $p$-spin model.}\BibitemShut
  {Stop}%
\bibitem [{\citenamefont {Derrida}(1980)}]{Derrida1980Random}%
  \BibitemOpen
  \bibfield  {author} {\bibinfo {author} {\bibfnamefont {B.}~\bibnamefont
  {Derrida}},\ }\href@noop {} {\bibfield  {journal} {\bibinfo  {journal} {Phys.
  Rev. Lett.}\ }\textbf {\bibinfo {volume} {45}},\ \bibinfo {pages} {79}
  (\bibinfo {year} {1980})}\BibitemShut {NoStop}%
\bibitem [{\citenamefont {Gross}\ and\ \citenamefont
  {Mezard}(1984)}]{Gross1984Simplest}%
  \BibitemOpen
  \bibfield  {author} {\bibinfo {author} {\bibfnamefont {D.}~\bibnamefont
  {Gross}}\ and\ \bibinfo {author} {\bibfnamefont {M.}~\bibnamefont {Mezard}},\
  }\href@noop {} {\bibfield  {journal} {\bibinfo  {journal} {Nuclear Physics
  B}\ }\textbf {\bibinfo {volume} {240}},\ \bibinfo {pages} {431 } (\bibinfo
  {year} {1984})}\BibitemShut {NoStop}%
\bibitem [{\citenamefont {Gardner}(1985)}]{Gardner1985Spin}%
  \BibitemOpen
  \bibfield  {author} {\bibinfo {author} {\bibfnamefont {E.}~\bibnamefont
  {Gardner}},\ }\href@noop {} {\bibfield  {journal} {\bibinfo  {journal}
  {Nuclear Physics B}\ }\textbf {\bibinfo {volume} {257}},\ \bibinfo {pages}
  {747 } (\bibinfo {year} {1985})}\BibitemShut {NoStop}%
\bibitem [{\citenamefont {Kirkpatrick}\ and\ \citenamefont
  {Thirumalai}(1987)}]{Kirkpatrick1987pSpin}%
  \BibitemOpen
  \bibfield  {author} {\bibinfo {author} {\bibfnamefont {T.~R.}\ \bibnamefont
  {Kirkpatrick}}\ and\ \bibinfo {author} {\bibfnamefont {D.}~\bibnamefont
  {Thirumalai}},\ }\href@noop {} {\bibfield  {journal} {\bibinfo  {journal}
  {Phys. Rev. B}\ }\textbf {\bibinfo {volume} {36}},\ \bibinfo {pages} {5388}
  (\bibinfo {year} {1987})}\BibitemShut {NoStop}%
\bibitem [{\citenamefont {Biroli}\ and\ \citenamefont
  {Bouchaud}(2012)}]{Biroli2012Random}%
  \BibitemOpen
  \bibfield  {author} {\bibinfo {author} {\bibfnamefont {G.}~\bibnamefont
  {Biroli}}\ and\ \bibinfo {author} {\bibfnamefont {J.~P.}\ \bibnamefont
  {Bouchaud}},\ }\href@noop {} {\emph {\bibinfo {title} {The Random First-Order
  Transition Theory of Glasses}}},\ edited by\ \bibinfo {editor} {\bibfnamefont
  {P.~G.}\ \bibnamefont {Wolynes}}\ and\ \bibinfo {editor} {\bibfnamefont
  {V.}~\bibnamefont {Lubchenko}}\ (\bibinfo  {publisher} {John Wiley \& Sons,
  Inc.},\ \bibinfo {year} {2012})\ Chap.~\bibinfo {chapter} {2}\BibitemShut
  {NoStop}%
\bibitem [{\citenamefont {M\'ezard}\ and\ \citenamefont
  {Montanari}(2009)}]{Mezard2009}%
  \BibitemOpen
  \bibfield  {author} {\bibinfo {author} {\bibfnamefont {M.}~\bibnamefont
  {M\'ezard}}\ and\ \bibinfo {author} {\bibfnamefont {A.}~\bibnamefont
  {Montanari}},\ }\href@noop {} {\emph {\bibinfo {title} {Information, Physics,
  and Computation}}},\ Oxford Graduate Texts\ (\bibinfo  {publisher} {OUP
  Oxford},\ \bibinfo {year} {2009})\BibitemShut {NoStop}%
\bibitem [{\citenamefont {Bapst}\ \emph {et~al.}(2013)\citenamefont {Bapst},
  \citenamefont {Foini}, \citenamefont {Krzakala}, \citenamefont {Semerjian},\
  and\ \citenamefont {Zamponi}}]{Bapst2013Quantum}%
  \BibitemOpen
  \bibfield  {author} {\bibinfo {author} {\bibfnamefont {V.}~\bibnamefont
  {Bapst}}, \bibinfo {author} {\bibfnamefont {L.}~\bibnamefont {Foini}},
  \bibinfo {author} {\bibfnamefont {F.}~\bibnamefont {Krzakala}}, \bibinfo
  {author} {\bibfnamefont {G.}~\bibnamefont {Semerjian}}, \ and\ \bibinfo
  {author} {\bibfnamefont {F.}~\bibnamefont {Zamponi}},\ }\href@noop {}
  {\bibfield  {journal} {\bibinfo  {journal} {Physics Reports}\ }\textbf
  {\bibinfo {volume} {523}},\ \bibinfo {pages} {127 } (\bibinfo {year}
  {2013})}\BibitemShut {NoStop}%
\bibitem [{\citenamefont {{J. Kurchan}}\ \emph {et~al.}(1993)\citenamefont {{J.
  Kurchan}}, \citenamefont {{G.Parisi}},\ and\ \citenamefont {{M.A.
  Virasoro}}}]{Kurchan1993Barriers}%
  \BibitemOpen
  \bibfield  {author} {\bibinfo {author} {\bibnamefont {{J. Kurchan}}},
  \bibinfo {author} {\bibnamefont {{G.Parisi}}}, \ and\ \bibinfo {author}
  {\bibnamefont {{M.A. Virasoro}}},\ }\href@noop {} {\bibfield  {journal}
  {\bibinfo  {journal} {J. Phys. I France}\ }\textbf {\bibinfo {volume} {3}},\
  \bibinfo {pages} {1819} (\bibinfo {year} {1993})}\BibitemShut {NoStop}%
\bibitem [{\citenamefont {{Silvio Franz}}\ and\ \citenamefont {{Giorgio
  Parisi}}(1995)}]{Franz1995Recipes}%
  \BibitemOpen
  \bibfield  {author} {\bibinfo {author} {\bibnamefont {{Silvio Franz}}}\ and\
  \bibinfo {author} {\bibnamefont {{Giorgio Parisi}}},\ }\href@noop {}
  {\bibfield  {journal} {\bibinfo  {journal} {J. Phys. I France}\ }\textbf
  {\bibinfo {volume} {5}},\ \bibinfo {pages} {1401} (\bibinfo {year}
  {1995})}\BibitemShut {NoStop}%
\bibitem [{\citenamefont {Franz}\ and\ \citenamefont
  {Parisi}(1997)}]{Franz1997Phase}%
  \BibitemOpen
  \bibfield  {author} {\bibinfo {author} {\bibfnamefont {S.}~\bibnamefont
  {Franz}}\ and\ \bibinfo {author} {\bibfnamefont {G.}~\bibnamefont {Parisi}},\
  }\href@noop {} {\bibfield  {journal} {\bibinfo  {journal} {Phys. Rev. Lett.}\
  }\textbf {\bibinfo {volume} {79}},\ \bibinfo {pages} {2486} (\bibinfo {year}
  {1997})}\BibitemShut {NoStop}%
\bibitem [{\citenamefont {Sompolinsky}\ and\ \citenamefont
  {Zippelius}(1981)}]{Sompolinsky1981Dynamic}%
  \BibitemOpen
  \bibfield  {author} {\bibinfo {author} {\bibfnamefont {H.}~\bibnamefont
  {Sompolinsky}}\ and\ \bibinfo {author} {\bibfnamefont {A.}~\bibnamefont
  {Zippelius}},\ }\href@noop {} {\bibfield  {journal} {\bibinfo  {journal}
  {Phys. Rev. Lett.}\ }\textbf {\bibinfo {volume} {47}},\ \bibinfo {pages}
  {359} (\bibinfo {year} {1981})}\BibitemShut {NoStop}%
\bibitem [{\citenamefont {Anderson}(1958)}]{Anderson1958Absence}%
  \BibitemOpen
  \bibfield  {author} {\bibinfo {author} {\bibfnamefont {P.~W.}\ \bibnamefont
  {Anderson}},\ }\href@noop {} {\bibfield  {journal} {\bibinfo  {journal}
  {Phys. Rev.}\ }\textbf {\bibinfo {volume} {109}},\ \bibinfo {pages} {1492}
  (\bibinfo {year} {1958})}\BibitemShut {NoStop}%
\bibitem [{\citenamefont {Thouless}(1974)}]{Thouless1974Electrons}%
  \BibitemOpen
  \bibfield  {author} {\bibinfo {author} {\bibfnamefont {D.}~\bibnamefont
  {Thouless}},\ }\href@noop {} {\bibfield  {journal} {\bibinfo  {journal}
  {Physics Reports}\ }\textbf {\bibinfo {volume} {13}},\ \bibinfo {pages} {93 }
  (\bibinfo {year} {1974})}\BibitemShut {NoStop}%
\bibitem [{\citenamefont {Bravyi}\ \emph {et~al.}(2011)\citenamefont {Bravyi},
  \citenamefont {DiVincenzo},\ and\ \citenamefont
  {Loss}}]{Bravyi2011Schrieffer}%
  \BibitemOpen
  \bibfield  {author} {\bibinfo {author} {\bibfnamefont {S.}~\bibnamefont
  {Bravyi}}, \bibinfo {author} {\bibfnamefont {D.~P.}\ \bibnamefont
  {DiVincenzo}}, \ and\ \bibinfo {author} {\bibfnamefont {D.}~\bibnamefont
  {Loss}},\ }\href@noop {} {\bibfield  {journal} {\bibinfo  {journal} {Annals
  of Physics}\ }\textbf {\bibinfo {volume} {326}},\ \bibinfo {pages} {2793 }
  (\bibinfo {year} {2011})}\BibitemShut {NoStop}%
\bibitem [{\citenamefont {Pietracaprina}\ \emph {et~al.}(2016)\citenamefont
  {Pietracaprina}, \citenamefont {Ros},\ and\ \citenamefont
  {Scardicchio}}]{Pietracaprina2016Forward}%
  \BibitemOpen
  \bibfield  {author} {\bibinfo {author} {\bibfnamefont {F.}~\bibnamefont
  {Pietracaprina}}, \bibinfo {author} {\bibfnamefont {V.}~\bibnamefont {Ros}},
  \ and\ \bibinfo {author} {\bibfnamefont {A.}~\bibnamefont {Scardicchio}},\
  }\href@noop {} {\bibfield  {journal} {\bibinfo  {journal} {Phys. Rev. B}\
  }\textbf {\bibinfo {volume} {93}},\ \bibinfo {pages} {054201} (\bibinfo
  {year} {2016})}\BibitemShut {NoStop}%
\bibitem [{\citenamefont {J\"org}\ \emph {et~al.}(2008)\citenamefont {J\"org},
  \citenamefont {Krzakala}, \citenamefont {Kurchan},\ and\ \citenamefont
  {Maggs}}]{Jorg2008Simple}%
  \BibitemOpen
  \bibfield  {author} {\bibinfo {author} {\bibfnamefont {T.}~\bibnamefont
  {J\"org}}, \bibinfo {author} {\bibfnamefont {F.}~\bibnamefont {Krzakala}},
  \bibinfo {author} {\bibfnamefont {J.}~\bibnamefont {Kurchan}}, \ and\
  \bibinfo {author} {\bibfnamefont {A.~C.}\ \bibnamefont {Maggs}},\ }\href@noop
  {} {\bibfield  {journal} {\bibinfo  {journal} {Phys. Rev. Lett.}\ }\textbf
  {\bibinfo {volume} {101}},\ \bibinfo {pages} {147204} (\bibinfo {year}
  {2008})}\BibitemShut {NoStop}%
\bibitem [{\citenamefont {Sakurai}(2011)}]{Sakurai2011}%
  \BibitemOpen
  \bibfield  {author} {\bibinfo {author} {\bibfnamefont {J.~J.}\ \bibnamefont
  {Sakurai}},\ }\href@noop {} {\emph {\bibinfo {title} {Modern Quantum
  Mechanics}}}\ (\bibinfo  {publisher} {Addison-Wesley},\ \bibinfo {year}
  {2011})\BibitemShut {NoStop}%
\bibitem [{\citenamefont {Laumann}\ \emph {et~al.}(2014)\citenamefont
  {Laumann}, \citenamefont {Pal},\ and\ \citenamefont
  {Scardicchio}}]{Laumann2014Many}%
  \BibitemOpen
  \bibfield  {author} {\bibinfo {author} {\bibfnamefont {C.~R.}\ \bibnamefont
  {Laumann}}, \bibinfo {author} {\bibfnamefont {A.}~\bibnamefont {Pal}}, \ and\
  \bibinfo {author} {\bibfnamefont {A.}~\bibnamefont {Scardicchio}},\
  }\href@noop {} {\bibfield  {journal} {\bibinfo  {journal} {Phys. Rev. Lett.}\
  }\textbf {\bibinfo {volume} {113}},\ \bibinfo {pages} {200405} (\bibinfo
  {year} {2014})}\BibitemShut {NoStop}%
\bibitem [{\citenamefont {Goldschmidt}(1990)}]{Goldschmidt1990Solvable}%
  \BibitemOpen
  \bibfield  {author} {\bibinfo {author} {\bibfnamefont {Y.~Y.}\ \bibnamefont
  {Goldschmidt}},\ }\href@noop {} {\bibfield  {journal} {\bibinfo  {journal}
  {Phys. Rev. B}\ }\textbf {\bibinfo {volume} {41}},\ \bibinfo {pages} {4858}
  (\bibinfo {year} {1990})}\BibitemShut {NoStop}%
\bibitem [{Note2()}]{Note2}%
  \BibitemOpen
  \bibinfo {note} {To obtain Eq.~\protect \textup {\hbox {\mathsurround \z@
  \protect \normalfont (\ignorespaces \ref
  {eq:effective_coupling_large_p_expansion}\unskip \@@italiccorr )}} from
  Eq.~\protect \textup {\hbox {\mathsurround \z@ \protect \normalfont
  (\ignorespaces \ref {eq:finite_p_effective_coupling_exponent}\unskip
  \@@italiccorr )}}, substitute $y = w/p$ and take $\mu (w/p) \sim
  e^{-2w}$.}\BibitemShut {Stop}%
\bibitem [{\citenamefont {Dobrosavljevic}\ and\ \citenamefont
  {Thirumalai}(1990)}]{Dobrosavljevic1990Expansion}%
  \BibitemOpen
  \bibfield  {author} {\bibinfo {author} {\bibfnamefont {V.}~\bibnamefont
  {Dobrosavljevic}}\ and\ \bibinfo {author} {\bibfnamefont {D.}~\bibnamefont
  {Thirumalai}},\ }\href@noop {} {\bibfield  {journal} {\bibinfo  {journal}
  {Journal of Physics A: Mathematical and General}\ }\textbf {\bibinfo {volume}
  {23}},\ \bibinfo {pages} {L767} (\bibinfo {year} {1990})}\BibitemShut
  {NoStop}%
\bibitem [{\citenamefont {Nieuwenhuizen}\ and\ \citenamefont
  {Ritort}(1998)}]{Nieuwenhuizen19988}%
  \BibitemOpen
  \bibfield  {author} {\bibinfo {author} {\bibfnamefont {T.~M.}\ \bibnamefont
  {Nieuwenhuizen}}\ and\ \bibinfo {author} {\bibfnamefont {F.}~\bibnamefont
  {Ritort}},\ }\href@noop {} {\bibfield  {journal} {\bibinfo  {journal}
  {Physica A: Statistical Mechanics and its Applications}\ }\textbf {\bibinfo
  {volume} {250}},\ \bibinfo {pages} {8 } (\bibinfo {year} {1998})}\BibitemShut
  {NoStop}%
\bibitem [{\citenamefont {Biroli}\ \emph {et~al.}(2000)\citenamefont {Biroli},
  \citenamefont {Monasson},\ and\ \citenamefont
  {Weigt}}]{Biroli2000Variational}%
  \BibitemOpen
  \bibfield  {author} {\bibinfo {author} {\bibfnamefont {G.}~\bibnamefont
  {Biroli}}, \bibinfo {author} {\bibfnamefont {R.}~\bibnamefont {Monasson}}, \
  and\ \bibinfo {author} {\bibfnamefont {M.}~\bibnamefont {Weigt}},\
  }\href@noop {} {\bibfield  {journal} {\bibinfo  {journal} {The European
  Physical Journal B - Condensed Matter and Complex Systems}\ }\textbf
  {\bibinfo {volume} {14}},\ \bibinfo {pages} {551} (\bibinfo {year}
  {2000})}\BibitemShut {NoStop}%
\bibitem [{\citenamefont {M\'ezard}\ and\ \citenamefont
  {Zecchina}(2002)}]{Mezard2002Random}%
  \BibitemOpen
  \bibfield  {author} {\bibinfo {author} {\bibfnamefont {M.}~\bibnamefont
  {M\'ezard}}\ and\ \bibinfo {author} {\bibfnamefont {R.}~\bibnamefont
  {Zecchina}},\ }\href@noop {} {\bibfield  {journal} {\bibinfo  {journal}
  {Phys. Rev. E}\ }\textbf {\bibinfo {volume} {66}},\ \bibinfo {pages} {056126}
  (\bibinfo {year} {2002})}\BibitemShut {NoStop}%
\bibitem [{\citenamefont {M\'ezard}\ \emph {et~al.}(2003)\citenamefont
  {M\'ezard}, \citenamefont {Ricci-Tersenghi},\ and\ \citenamefont
  {Zecchina}}]{Mezard2003Two}%
  \BibitemOpen
  \bibfield  {author} {\bibinfo {author} {\bibfnamefont {M.}~\bibnamefont
  {M\'ezard}}, \bibinfo {author} {\bibfnamefont {F.}~\bibnamefont
  {Ricci-Tersenghi}}, \ and\ \bibinfo {author} {\bibfnamefont {R.}~\bibnamefont
  {Zecchina}},\ }\href@noop {} {\bibfield  {journal} {\bibinfo  {journal}
  {Journal of Statistical Physics}\ }\textbf {\bibinfo {volume} {111}},\
  \bibinfo {pages} {505} (\bibinfo {year} {2003})}\BibitemShut {NoStop}%
\bibitem [{\citenamefont {Zdeborov\'a}\ and\ \citenamefont
  {Krzakala}(2007)}]{Zdeborova2007Phase}%
  \BibitemOpen
  \bibfield  {author} {\bibinfo {author} {\bibfnamefont {L.}~\bibnamefont
  {Zdeborov\'a}}\ and\ \bibinfo {author} {\bibfnamefont {F.}~\bibnamefont
  {Krzakala}},\ }\href@noop {} {\bibfield  {journal} {\bibinfo  {journal}
  {Phys. Rev. E}\ }\textbf {\bibinfo {volume} {76}},\ \bibinfo {pages} {031131}
  (\bibinfo {year} {2007})}\BibitemShut {NoStop}%
\bibitem [{\citenamefont {Kirkpatrick}\ and\ \citenamefont
  {Selman}(1994)}]{Kirkpatrick1994Critical}%
  \BibitemOpen
  \bibfield  {author} {\bibinfo {author} {\bibfnamefont {S.}~\bibnamefont
  {Kirkpatrick}}\ and\ \bibinfo {author} {\bibfnamefont {B.}~\bibnamefont
  {Selman}},\ }\href@noop {} {\bibfield  {journal} {\bibinfo  {journal}
  {Science}\ }\textbf {\bibinfo {volume} {264}},\ \bibinfo {pages} {1297}
  (\bibinfo {year} {1994})}\BibitemShut {NoStop}%
\bibitem [{\citenamefont {Farhi}\ \emph {et~al.}(2008)\citenamefont {Farhi},
  \citenamefont {Goldstone}, \citenamefont {Gutmann},\ and\ \citenamefont
  {Nagaj}}]{Farhi2008How}%
  \BibitemOpen
  \bibfield  {author} {\bibinfo {author} {\bibfnamefont {E.}~\bibnamefont
  {Farhi}}, \bibinfo {author} {\bibfnamefont {J.}~\bibnamefont {Goldstone}},
  \bibinfo {author} {\bibfnamefont {S.}~\bibnamefont {Gutmann}}, \ and\
  \bibinfo {author} {\bibfnamefont {D.}~\bibnamefont {Nagaj}},\ }\href@noop {}
  {\bibfield  {journal} {\bibinfo  {journal} {International Journal of Quantum
  Information}\ }\textbf {\bibinfo {volume} {6}},\ \bibinfo {pages} {503}
  (\bibinfo {year} {2008})}\BibitemShut {NoStop}%
\bibitem [{\citenamefont {Nielsen}\ and\ \citenamefont
  {Chuang}(2000)}]{Nielsen2000Quantum}%
  \BibitemOpen
  \bibfield  {author} {\bibinfo {author} {\bibfnamefont {M.~A.}\ \bibnamefont
  {Nielsen}}\ and\ \bibinfo {author} {\bibfnamefont {I.~L.}\ \bibnamefont
  {Chuang}},\ }\href@noop {} {\emph {\bibinfo {title} {Quantum Computation and
  Quantum Information}}}\ (\bibinfo  {publisher} {Cambridge University Press},\
  \bibinfo {year} {2000})\BibitemShut {NoStop}%
\bibitem [{\citenamefont {Johnson}\ \emph {et~al.}(2011)\citenamefont
  {Johnson}, \citenamefont {Amin}, \citenamefont {Gildert}, \citenamefont
  {Lanting}, \citenamefont {Hamze}, \citenamefont {Dickson}, \citenamefont
  {Harris}, \citenamefont {Berkley}, \citenamefont {Johansson}, \citenamefont
  {Bunyk}, \citenamefont {Chapple}, \citenamefont {Enderud}, \citenamefont
  {Hilton}, \citenamefont {Karimi}, \citenamefont {Ladizinsky}, \citenamefont
  {Ladizinsky}, \citenamefont {Oh}, \citenamefont {Perminov}, \citenamefont
  {Rich}, \citenamefont {Thom}, \citenamefont {Tolkacheva}, \citenamefont
  {Truncik}, \citenamefont {Uchaikin}, \citenamefont {Wang}, \citenamefont
  {Wilson},\ and\ \citenamefont {Rose}}]{Johnson2011Quantum}%
  \BibitemOpen
  \bibfield  {author} {\bibinfo {author} {\bibfnamefont {M.~W.}\ \bibnamefont
  {Johnson}}, \bibinfo {author} {\bibfnamefont {M.~H.~S.}\ \bibnamefont
  {Amin}}, \bibinfo {author} {\bibfnamefont {S.}~\bibnamefont {Gildert}},
  \bibinfo {author} {\bibfnamefont {T.}~\bibnamefont {Lanting}}, \bibinfo
  {author} {\bibfnamefont {F.}~\bibnamefont {Hamze}}, \bibinfo {author}
  {\bibfnamefont {N.}~\bibnamefont {Dickson}}, \bibinfo {author} {\bibfnamefont
  {R.}~\bibnamefont {Harris}}, \bibinfo {author} {\bibfnamefont {A.~J.}\
  \bibnamefont {Berkley}}, \bibinfo {author} {\bibfnamefont {J.}~\bibnamefont
  {Johansson}}, \bibinfo {author} {\bibfnamefont {P.}~\bibnamefont {Bunyk}},
  \bibinfo {author} {\bibfnamefont {E.~M.}\ \bibnamefont {Chapple}}, \bibinfo
  {author} {\bibfnamefont {C.}~\bibnamefont {Enderud}}, \bibinfo {author}
  {\bibfnamefont {J.~P.}\ \bibnamefont {Hilton}}, \bibinfo {author}
  {\bibfnamefont {K.}~\bibnamefont {Karimi}}, \bibinfo {author} {\bibfnamefont
  {E.}~\bibnamefont {Ladizinsky}}, \bibinfo {author} {\bibfnamefont
  {N.}~\bibnamefont {Ladizinsky}}, \bibinfo {author} {\bibfnamefont
  {T.}~\bibnamefont {Oh}}, \bibinfo {author} {\bibfnamefont {I.}~\bibnamefont
  {Perminov}}, \bibinfo {author} {\bibfnamefont {C.}~\bibnamefont {Rich}},
  \bibinfo {author} {\bibfnamefont {M.~C.}\ \bibnamefont {Thom}}, \bibinfo
  {author} {\bibfnamefont {E.}~\bibnamefont {Tolkacheva}}, \bibinfo {author}
  {\bibfnamefont {C.~J.~S.}\ \bibnamefont {Truncik}}, \bibinfo {author}
  {\bibfnamefont {S.}~\bibnamefont {Uchaikin}}, \bibinfo {author}
  {\bibfnamefont {J.}~\bibnamefont {Wang}}, \bibinfo {author} {\bibfnamefont
  {B.}~\bibnamefont {Wilson}}, \ and\ \bibinfo {author} {\bibfnamefont
  {G.}~\bibnamefont {Rose}},\ }\href@noop {} {\bibfield  {journal} {\bibinfo
  {journal} {Nature}\ }\textbf {\bibinfo {volume} {473}},\ \bibinfo {pages}
  {194} (\bibinfo {year} {2011})}\BibitemShut {NoStop}%
\bibitem [{\citenamefont {Boixo}\ \emph {et~al.}(2014)\citenamefont {Boixo},
  \citenamefont {Ronnow}, \citenamefont {Isakov}, \citenamefont {Wang},
  \citenamefont {Wecker}, \citenamefont {Lidar}, \citenamefont {Martinis},\
  and\ \citenamefont {Troyer}}]{Boixo2014Evidence}%
  \BibitemOpen
  \bibfield  {author} {\bibinfo {author} {\bibfnamefont {S.}~\bibnamefont
  {Boixo}}, \bibinfo {author} {\bibfnamefont {T.~F.}\ \bibnamefont {Ronnow}},
  \bibinfo {author} {\bibfnamefont {S.~V.}\ \bibnamefont {Isakov}}, \bibinfo
  {author} {\bibfnamefont {Z.}~\bibnamefont {Wang}}, \bibinfo {author}
  {\bibfnamefont {D.}~\bibnamefont {Wecker}}, \bibinfo {author} {\bibfnamefont
  {D.~A.}\ \bibnamefont {Lidar}}, \bibinfo {author} {\bibfnamefont {J.~M.}\
  \bibnamefont {Martinis}}, \ and\ \bibinfo {author} {\bibfnamefont
  {M.}~\bibnamefont {Troyer}},\ }\href@noop {} {\bibfield  {journal} {\bibinfo
  {journal} {Nat Phys}\ }\textbf {\bibinfo {volume} {10}},\ \bibinfo {pages}
  {218} (\bibinfo {year} {2014})}\BibitemShut {NoStop}%
\bibitem [{Note3()}]{Note3}%
  \BibitemOpen
  \bibinfo {note} {It would be interesting to show that matching finite energy
  density solutions is likewise NP-complete, analogous to a hardness of
  approximation result, although we have not done so.}\BibitemShut {Stop}%
\bibitem [{Note4()}]{Note4}%
  \BibitemOpen
  \bibinfo {note} {By Markov's inequality, a typical sample cannot have more
  than a finite factor times the expected number of paths as $N \rightarrow
  \infty $.}\BibitemShut {Stop}%
\end{thebibliography}%

\appendix

\section{NP-completeness of ground-state energy matching in K-CSPs} \label{appendix:np_complete}

Here we show that the matching problem for the ground states of $K$-body constraint satisfaction problems (CSPs) is NP-complete when $K \geq 4$. 
To be precise, we define the decision problem~\footnote{It would be interesting to show that matching finite energy density solutions is likewise NP-complete, analogous to a hardness of approximation result, although we have not done so.}:

\vspace{1mm}
\fbox{
\begin{minipage}[t]{3in}
{\bf \underline{Matching-K-CSP}} \vspace{2mm} \\
\textbf{Input: } $(H, \sigma, x)$, where $H$ is a classical energy function encoding a $K$-CSP, $\sigma \in \{0,1\}^N$ is a satisfying ground state of $H$ and $x \in (0,1)$ is a fractional Hamming distance. \\
\textbf{Output: } YES if there exists a satisfying state $\sigma'$ such that $x(\sigma, \sigma') \ge x$.
\end{minipage}
}
\vspace{1mm}

It is clear that matching-$K$-CSP is in NP.
In order to show that it is NP-complete, we construct a reduction from $(K-1)$-CSP to matching-$K$-CSP.
Since $(K-1)$-CSP is NP-complete for $K-1 \geq 3$, this proves that matching-$K$-CSP is NP-complete for $K \geq 4$.

Let $H_p$ be an instance of $(K-1)$-CSP acting on $N$ bits $s_i$. 
We construct an instance of matching-$K$-CSP by introducing $M = 2 N$ auxiliary bits $\tau_j$ and defining the $K$-CSP Hamiltonian:
\begin{equation} \label{eq:matching_Hamiltonian}
\begin{aligned}
\tilde{H}_p =& \; H_p \otimes \left( \frac{1}{2M} \sum_{j=1}^M \big( 1 + \tau_j \big) \right) \\
& \qquad \qquad + I \otimes \left( M - \frac{1}{M} \sum_{j,j'=1}^{M} \tau_j \tau_{j'} \right) ,
\end{aligned}
\end{equation}
where the left factor in each tensor product refers to the original bits and the right factor to the auxiliary bits.
The first term interpolates between 0 when all $\tau_j = -1$ and $H_p$ when all $\tau_j = +1$.
The second term contributes an energy cost when any two auxiliary bits are anti-aligned.
Clearly $\tilde{H}_p$ involves no more than $K$-body interactions.

A state $\ket{s} \otimes \ket{\tau}$ has zero energy under $\tilde{H}_p$ if it satisfies \textit{one of} the following conditions:
\begin{enumerate}
	\item All $\tau_j = -1$.
	\item All $\tau_j = +1$ and $H_p \ket{s} = 0$.
\end{enumerate}
All other states have positive energy under $\tilde{H}_p$.
The given zero-energy state of the matching problem is $\ket{-} \otimes \ket{-}$, i.e., all original and auxiliary bits have value $-1$, which satisfies condition 1.

It is then clear that $H_p$ has a zero-energy state if and only if $\tilde{H}_p$ has a zero-energy state at distance greater than 1/3 from $\ket{-} \otimes \ket{-}$.
If $H_p$ has a zero-energy state $\ket{s}$, then $\tilde{H}_p \big( \ket{s} \otimes \ket{+} \big) = 0$ and $\ket{s} \otimes \ket{+}$ is at distance greater than 2/3 from $\ket{-} \otimes \ket{-}$.
If $\tilde{H}_p$ has a zero-energy state $\ket{s} \otimes \ket{\tau}$ at sufficient distance, then $\ket{s} \otimes \ket{\tau}$ cannot satisfy condition 1 as that would imply a distance less than 1/3 and so must satisfy condition 2, i.e., $\ket{s}$ is a satisfying configuration of $H_p$.
We thus have the desired mapping from any instance $H_p$ of $(K-1)$-CSP to an instance $(\tilde{H}_p, \ket{-} \otimes \ket{-}, 1/3)$ of matching-$K$-CSP.

\section{The FPP in the canonical and microcanonical ensembles} \label{appendix:fpp}

The Franz-Parisi potential is typically presented in the canonical ensemble~\cite{Kurchan1993Barriers,Franz1995Recipes,Franz1997Phase,Mezard2009}.
It represents the free energy of a system at inverse temperature $\beta'$ constrained to be at distance $x$ from a system in equilibrium at inverse temperature $\beta$:
\begin{widetext}
\begin{equation} \label{eq:standard_fpp}
v(x, \beta' | \beta) = \mathbb{E} \Bigg[ \textrm{Tr}_{\sigma} \left[ \frac{e^{-N \beta \epsilon(\sigma)}}{Z(\beta)} \ln{\textrm{Tr}_{\sigma'} \big[ \delta_{x, x(\sigma, \sigma')} e^{-N \beta' \epsilon(\sigma')} \big] } \right] \Bigg] ,
\end{equation}
\end{widetext}
where $Z(\beta)$ is the partition function.
However, in the present paper it is more natural to use $g(x, \epsilon' | \epsilon)$ as defined in Eq.~\eqref{eq:microcanonical_fpp}.
Here we show the relationship between the two.

First consider
\begin{equation*}
\mathbb{E} \Bigg[ \textrm{Tr}_{\sigma} \left[ \frac{e^{-N \beta \epsilon(\sigma)}}{Z(\beta)} f(\sigma) \right] \Bigg] ,
\end{equation*}
for a function $f(\sigma)$ growing slower than exponential with $N$ but otherwise arbitrary.
The trace is dominated by those $\sigma$ at the energy density $\epsilon$ which maximizes $s(\epsilon) - \beta \epsilon$, where $s(\epsilon)$ is the disorder-averaged entropy density (the sample-to-sample fluctuations about $s(\epsilon)$ are expected to vanish as $N \rightarrow \infty$).
The configurations at other energy densities collectively give an exponentially small contribution.
Thus as $N \rightarrow \infty$,
\begin{equation} \label{eq:general_boltzmann_weighting}
\mathbb{E} \Bigg[ \textrm{Tr}_{\sigma} \left[ \frac{e^{-N \beta \epsilon(\sigma)}}{Z(\beta)} f(\sigma) \right] \Bigg] \sim \mathbb{E} \Bigg[ \textrm{Tr}_{\sigma} \left[ \frac{\delta \big( \epsilon - \epsilon(\sigma) \big) }{\mathcal{N}(\epsilon)} f(\sigma) \right] \Bigg] ,
\end{equation}
where $\epsilon = \argmax_z \big[ s(z) - \beta z \big]$ and $\mathcal{N}(\epsilon) = e^{Ns(\epsilon)}$.

Eq.~\eqref{eq:general_boltzmann_weighting} applies to $v(x, \beta' | \beta)$ with
\begin{equation} \label{eq:standard_fpp_summand}
f(\sigma) = \ln{\textrm{Tr}_{\sigma'} \big[ \delta_{x, x(\sigma, \sigma')} e^{-N \beta' \epsilon(\sigma')} \big] }.
\end{equation}
In fact, since all configurations $\sigma$ are statistically equivalent, the argument of the trace is independent of $\sigma$ and we have that
\begin{equation} \label{eq:intermediate_fpp}
v(x, \beta' | \beta) = \mathbb{E} \Big[ \ln{\textrm{Tr}_{\sigma'} \big[ \delta_{x, x(\sigma, \sigma')} e^{-N \beta' \epsilon(\sigma')} \big] } \Big] \bigg|_{\epsilon(\sigma) = \epsilon} ,
\end{equation}
for an arbitrary reference configuration $\sigma$.

Next write
\begin{equation} \label{eq:boltzmann_micro_connection}
\begin{aligned}
& \textrm{Tr}_{\sigma'} \big[ \delta_{x, x(\sigma, \sigma')} e^{-N \beta' \epsilon(\sigma')} \big] \\
& \qquad = \int \textrm{d}\epsilon' e^{-N \beta' \epsilon'} \textrm{Tr}_{\sigma'} \big[ \delta_{x, x(\sigma, \sigma')} \delta \big( \epsilon' - \epsilon(\sigma') \big) \big] .
\end{aligned}
\end{equation}
Assuming, as is typical, that $g(x, \epsilon' | \epsilon)$ is self-averaging, the trace on the right-hand side is precisely $e^{N g(x, \epsilon' | \epsilon)}$ (cf. Eq.~\eqref{eq:microcanonical_fpp}).
The integral is dominated by $\epsilon' = \argmax_{z'} \big[ g(x, z' | \epsilon) - \beta' z' \big]$.
We thus have the desired relationship between the canonical FPP $v(x, \beta' | \beta)$ and the microcanonical FPP $g(x, \epsilon' | \epsilon)$;
\begin{equation} \label{eq:micro_canon_fpp}
v(x, \beta' | \beta) = g(x, \epsilon' | \epsilon) - \beta' \epsilon' ,
\end{equation}
\begin{equation} \label{eq:epsilon_determination}
\epsilon = \argmax_z \big[ s(z) - \beta z \big] ,
\end{equation}
\begin{equation} \label{eq:epsilon_prime_determination}
\epsilon' = \argmax_{z'} \big[ g(x, z' | \epsilon) - \beta' z' \big] .
\end{equation}
Note that the relationship between $\epsilon'$ and $\beta'$ is not set by the full entropy $s(\epsilon)$ but rather by the distance-resolved entropy $g(x, \epsilon' | \epsilon)$.

Eqs.~\eqref{eq:micro_canon_fpp},~\eqref{eq:epsilon_determination}, and~\eqref{eq:epsilon_prime_determination} express $v(x, \beta' | \beta)$ in terms of $g(x, \epsilon' | \epsilon)$ through a Legendre transform.
By inverting the transform, one obtains $g(x, \epsilon' | \epsilon)$ in terms of $v(x, \beta' | \beta)$.

\section{Correlations among energy levels} \label{appendix:energy_correlations}

Here we calculate the conditional expecation values needed in the main text, namely for obtaining $g(x, \epsilon' | \epsilon)$ and $\gamma(x, \epsilon' | \epsilon)$ at finite $p$ (Eqs.~\eqref{eq:microcanonical_fpp_result} and~\eqref{eq:finite_p_effective_coupling_exponent}).

The joint distribution of energy levels in the classical $p$-spin model follows immediately from the fact that the energies are Gaussian-distributed with covariances given by Eq.~\eqref{eq:p_spin_cov_matrix}.
For a subset of levels $\bm{\epsilon} \equiv \begin{pmatrix} \epsilon(\sigma_1) \; \cdots \; \epsilon(\sigma_n) \end{pmatrix} ^T$, the joint distribution is
\begin{equation} \label{eq:p_spin_joint_distribution}
P_n(\bm{\epsilon}) = e^{-N \bm{\epsilon}^T \Theta^{-1} \bm{\epsilon}},
\end{equation}
up to normalization, where
\begin{equation} \label{eq:distribution_corr_matrix}
\Theta \equiv
\begin{pmatrix}
1 & (1 - 2x_{12})^p & \cdots & (1 - 2x_{1n})^p \\
(1 - 2x_{12})^p & 1 & \cdots & (1 - 2x_{2n})^p \\
\vdots & \vdots & \ddots & \vdots \\
(1 - 2x_{1n})^p & (1 - 2x_{2n})^p & \cdots & 1
\end{pmatrix},
\end{equation}
with $x_{ij}$ shorthand for $x(\sigma_i, \sigma_j)$.
Thus the distribution for a single level $\epsilon$ is
\begin{equation} \label{eq:single_level_dist}
P_1(\epsilon) = e^{-N \epsilon^2},
\end{equation}
and the distribution for a pair of levels $\epsilon$ \& $\epsilon'$ is
\begin{equation} \label{eq:pair_level_dist}
P_2(\epsilon, \epsilon') = \exp{\left( -N \frac{\epsilon^2 - 2 (1 - 2x)^p \epsilon \epsilon' + \epsilon'^2}{1 - (1 - 2x)^{2p}} \right) }.
\end{equation}
The conditional distribution of $\epsilon'$ given $\epsilon$ is
\begin{equation} \label{eq:conditional_level_dist}
\frac{P_2(\epsilon', \epsilon)}{P_1(\epsilon)} = \exp{\left( -N \frac{\big( \epsilon' - (1 - 2x)^p \epsilon \big) ^2}{1 - (1 - 2x)^{2p}} \right) },
\end{equation}
from which Eq.~\eqref{eq:microcanonical_fpp_result} follows.

To obtain Eq.~\eqref{eq:intermediate_average_energy}, we need to average over $\epsilon''$ in the joint distribution $P_3(\epsilon, \epsilon', \epsilon'')$.
Write the 3 $\times$ 3 matrix $\Theta^{-1}$ in block-diagonal form as
\begin{equation} \label{eq:three_level_corr_matrix}
\Theta^{-1} = 
\begin{pmatrix}
A & \bm{b} \\
\bm{b}^T & c
\end{pmatrix},
\end{equation}
where $A$ is 2 $\times$ 2 and acts in the $(\epsilon, \epsilon')$ subspace, $\bm{b}$ is 2 $\times$ 1, and $c$ is 1 $\times$ 1.
Then
\begin{equation} \label{eq:three_level_dist}
\begin{aligned}
P_3(\epsilon, \epsilon', \epsilon'') =& \; \exp{\bigg( -N \big( c \epsilon''^2 + 2 \bm{b}^T \bm{\epsilon} \epsilon'' + \bm{\epsilon}^T A \bm{\epsilon} \big) \bigg)} \\
=& \; \exp{\bigg( -N c \Big( \epsilon'' + \frac{\bm{b}^T \bm{\epsilon}}{c} \Big) ^2 } \\
& \qquad \qquad \qquad - N \bm{\epsilon}^T \Big( A - \frac{\bm{b} \bm{b}^T}{c} \Big) \bm{\epsilon} \bigg) ,
\end{aligned}
\end{equation}
where $\bm{\epsilon} \equiv \begin{pmatrix} \epsilon \; \; \epsilon' \end{pmatrix} ^T$.
We see that
\begin{equation} \label{eq:conditional_third_energy_average}
\mathbb{E} \big[ \epsilon'' \big] \Big| _{\epsilon, \epsilon'} = - \frac{\bm{b}^T \bm{\epsilon}}{c}.
\end{equation}
A direct calculation of the inverse gives that
\begin{equation} \label{eq:three_matrix_inverse_coefs}
-\frac{1}{c} \bm{b} =
\begin{pmatrix}
\frac{(1 - 2y)^p - (1 - 2x)^p (1 - 2x + 2y)^p}{1 - (1 - 2x)^{2p}} \\[6pt]
\frac{(1 - 2x + 2y)^p - (1 - 2x)^p (1 - 2y)^p}{1 - (1 - 2x)^{2p}}
\end{pmatrix},
\end{equation}
from which Eqs.~\eqref{eq:intermediate_average_energy} and~\eqref{eq:finite_p_effective_coupling_exponent} follow.

\section{Schrieffer-Wolff and forward-scattering} \label{appendix:schrieffer_wolff}

\subsection{The effective coupling}

Consider a Hamiltonian $H = H_0 + \eta V$ with projectors $P_0$ and $Q_0 \equiv 1 - P_0$, such that $H_0 = P_0 H_0 P_0 + Q_0 H_0 Q_0$ (i.e., $H_0$ is block-diagonal with respect to $P_0$ and $Q_0$).
Owing to $V$, one may have that $H$ is not block-diagonal.
The Schrieffer-Wolff transformation is a unitary transformation $U$ such that $H_{\textrm{eff}} \equiv U H U^{\dag}$ \textit{is} block-diagonal, i.e., $P_0 H_{\textrm{eff}} Q_0 = Q_0 H_{\textrm{eff}} P_0 = 0$.
Here we discuss how to compute the generator of $U$ within the forward-scattering approximation, which amounts to retaining the lowest-order-in-$\eta$ terms for each matrix element of $H_{\textrm{eff}}$ (different matrix elements may become non-zero at different orders, and we work to lowest non-zero order for each separately).

We use the review by Bravyi et al~\cite{Bravyi2011Schrieffer} as our starting point.
First we present their notation.
$V$ is broken into diagonal and off-diagonal parts,
\begin{equation} \label{eq:diag_off_diag_decomposition}
\begin{aligned}
V_d & \equiv P_0 V P_0 + Q_0 V Q_0 , \\
V_{od} & \equiv P_0 V Q_0 + Q_0 V P_0 .
\end{aligned}
\end{equation}
Eigenstates of $H_0$ are used as the basis, denoted $\ket{i}, \ket{j}, \ldots$ with corresponding energies $E_i, E_j, \ldots$.
The unitary transformation $U$ is expressed as $e^{S}$, with $S$ \textit{anti}-Hermitian.
For consistency with~\cite{Bravyi2011Schrieffer}, we shall use this convention throughout the Appendix, even though we refer to a Hermitian generator in the main text.
Finally, Bravyi et al define superoperators
\begin{equation} \label{eq:adjoint_definition}
\hat{S} ( \, \cdot \, ) \equiv [ S \, , \, \cdot \, ] , 
\end{equation}
\begin{equation} \label{eq:L_definition}
\mathcal{L} ( \, \cdot \, ) \equiv \sum_{\substack{i \in P_0 \\ j \in Q_0}} \left( \ket{i} \frac{\braket{i | \cdot | j}}{E_i - E_j} \bra{j} + \ket{j} \frac{\braket{j | \cdot | i}}{E_j - E_i} \bra{i} \right) ,
\end{equation}
where $\cdot$ denotes an arbitrary operator.
Note that Eq.~\eqref{eq:L_definition} only has off-block-diagonal matrix elements.
With this notation, the condition that $H_{\textrm{eff}}$ be block-diagonal gives an equation for the generator $S$:
\begin{equation} \label{eq:block_diagonal_condition}
S = \mathcal{L} \hat{S} (\eta V_d) + \mathcal{L} \hat{S} \coth{(\hat{S})} (\eta V_{od}).
\end{equation}
The effective Hamiltonian is
\begin{equation} \label{eq:effective_Hamiltonian_expression}
H_{\textrm{eff}} = H_0 + \eta V_d + \tanh{\left( \frac{\hat{S}}{2} \right) } (\eta V_{od}).
\end{equation}
See~\cite{Bravyi2011Schrieffer} for the derivation.

Eq.~\eqref{eq:block_diagonal_condition} is naturally suited to an expansion in $\eta$:
\begin{equation} \label{eq:generator_expansion_form}
S = \sum_{n=1}^{\infty} \eta^n S_n,
\end{equation}
with each $S_n$ an anti-Hermitian operator.
From Eq.~\eqref{eq:block_diagonal_condition},
\begin{widetext}
\begin{equation} \label{eq:generator_expansion_orders}
\begin{aligned}
S_1 &= \mathcal{L}(V_{od}) , \\
S_2 &= \mathcal{L} \hat{S}_1 (V_d) , \\
S_n &= \mathcal{L} \hat{S}_{n-1} (V_d) \; + \; \sum_{j=1}^{\infty} a_{2j} \sum_{\substack{n_1, \cdots , n_{2j} \geq 1 \\ n_1 + \ldots + n_{2j} = n}} \mathcal{L} \hat{S}_{n_1} \cdots \hat{S}_{n_{2j}} (V_{od}) ,
\end{aligned}
\end{equation}
where the last line refers to $n \geq 3$, and $a_{2j}$ is the $2j$\textsuperscript{th} Taylor coefficient of $x \coth{x}$ about 0.
Consider the first two orders:
\begin{equation} \label{eq:generator_first_order}
S_1 = \sum_{\substack{i \in P_0 \\ j \in Q_0}} \left( \ket{i} \frac{\braket{i | V_{od} | j}}{E_i - E_j} \bra{j} + \ket{j} \frac{\braket{j | V_{od} | i}}{E_j - E_i} \bra{i} \right) ,
\end{equation}
\begin{equation} \label{eq:generator_second_order}
\begin{aligned}
S_2 =& \sum_{\substack{i \in P_0 \\ j \in Q_0 \\ k \in Q_0}} \left( \ket{i} \frac{\braket{i | V_{od} | j} \braket{j | V_d | k}}{(E_i - E_k)(E_i - E_j)} \bra{k} - \ket{k} \frac{\braket{k | V_d | j} \braket{j | V_{od} | i}}{(E_k - E_i)(E_j - E_i)} \bra{i} \right) \\
& \qquad \qquad \qquad + \sum_{\substack{i \in P_0 \\ j \in P_0 \\ k \in Q_0}} \left( \ket{i} \frac{\braket{i | V_d | j} \braket{j | V_{od} | k}}{(E_i - E_k)(E_j - E_k)} \bra{k} - \ket{k} \frac{\braket{k | V_{od} | j} \braket{j | V_d | i}}{(E_k - E_i)(E_k - E_j)} \bra{i} \right) .
\end{aligned}
\end{equation}
It becomes very tedious to write higher-order terms, yet one already sees the structure of the expansion.
Each term in $S_n$ has a numerator which is a string of matrix elements of $V$ and a denominator which is a string of energy differences.
Note that each matrix element can be either $V_d$ or $V_{od}$, and each energy denominator is between a state in $P_0$ and a state in $Q_0$.
The same structure holds when we insert the expansion of $S$ into Eq.~\eqref{eq:effective_Hamiltonian_expression} and obtain an expansion of $H_{\textrm{eff}}$ in powers of $\eta$.
For example, two of the fourth-order terms in $H_{\textrm{eff}}$ are
\begin{equation} \label{eq:effective_H_fourth_order_examples}
H_{\textrm{eff}} = \cdots + \frac{\eta^4}{2} \ket{i} \frac{\braket{i | V_{od} | j} \braket{j | V_d | k} \braket{k | V_d | l} \braket{l | V_{od} | m}}{(E_i - E_l)(E_i - E_k)(E_i - E_j)} \bra{m} + \frac{\eta^4}{2} \ket{i} \frac{\braket{i | V_d | j} \braket{j | V_{od} | k} \braket{k | V_d | l} \braket{l | V_{od} | m}}{(E_i - E_l)(E_i - E_k)(E_j - E_k)} \bra{m} + \cdots .
\end{equation}
\end{widetext}

Thus far, all that we have presented is completely general.
Now we show how the above equations, which are ultimately used to determine $H_{\textrm{eff}}$, are considerably simplified by making the forward-scattering approximation (FSA).
It is best to first consider the FSA within a simple toy problem, a 1D nearest-neighbor tight-binding model with open boundary conditions:
\begin{equation} \label{eq:toy_Hamiltonian}
H = \sum_{i=0}^L E_i \ket{i}\bra{i} - \eta \sum_{i=0}^{L-1} \Big( \ket{i}\bra{i+1} + \ket{i+1}\bra{i} \Big) .
\end{equation}
Suppose that $E_0$ and $E_L$ are much lower than all other $E_i$, and we want to study tunneling from site 0 to site $L$. We take $P_0$ to project onto $\ket{0}$ and $\ket{L}$, $V$ to be the hopping term, and aim to compute $\braket{0 | H_{\textrm{eff}} | L}$ to lowest order in $\eta$.

The lowest-order terms are $O \big( \eta^L \big)$, since at least $L$ applications of the hopping term are required to couple $\ket{0}$ and $\ket{L}$.
However, alongside those $L$\textsuperscript{th}-order terms which do couple $\ket{0}$ and $\ket{L}$, there are many $L$\textsuperscript{th}-order terms in $H_{\textrm{eff}}$ which do not.
Only terms for which the operator string is $V_{od} {V_d}^{L-2} V_{od}$ contribute to $\braket{0 | H_{\textrm{eff}} | L}$ at $L$\textsuperscript{th} order, since $\{ \ket{0}, \ket{L} \} \in P_0$ and $\{ \ket{1}, \ldots, \ket{L-1} \} \in Q_0$.
For example, taking $L = 4$, the first term in Eq.~\eqref{eq:effective_H_fourth_order_examples} does contribute to $\braket{0 | H_{\textrm{eff}} | 4}$ at 4\textsuperscript{th} order (taking $i = 0$, $j = 1$, etc.) but the second term does not.

In fact, only two terms at $L$\textsuperscript{th} order have the correct operator string:
\begin{widetext}
\begin{equation} \label{eq:toy_effective_coupling_v1}
\begin{aligned}
\braket{0 | H_{\textrm{eff}} | L} \sim & \frac{\eta^L}{2} \left( \frac{\braket{0 | V_{od} | 1} \braket{1 | V_d | 2} \cdots \braket{L-2 | V_d | L-1} \braket{L-1 | V_{od} | L}}{(E_0 - E_1)(E_0 - E_2) \cdots (E_0 - E_{L-2})(E_0 - E_{L-1})} \right. \\
& \qquad \qquad \qquad \qquad \qquad \qquad \left. + \frac{\braket{0 | V_{od} | 1} \braket{1 | V_d | 2} \cdots \braket{L-2 | V_d | L-1} \braket{L-1 | V_{od} | L}}{(E_L - E_1)(E_L - E_2) \cdots (E_L - E_{L-2})(E_L - E_{L-1})} \right) + O \big( \epsilon^{L+1} \big) .
\end{aligned}
\end{equation}
\end{widetext}
If $E_0 = E_L$, the final expression is particularly simple:
\begin{equation} \label{eq:toy_effective_coupling_v2}
\braket{0 | H_{\textrm{eff}} | L} \sim \eta \prod_{i=1}^{L-1} \frac{\eta}{E_0 - E_i} .
\end{equation}
Compare to Eq.~\eqref{eq:forward_scattering_expression} in the main text (noting that the toy problem has only one path from 0 to $L$).
This is the FSA, in which only the lowest-order terms in $\braket{0 | H_{\textrm{eff}} | L}$ are kept.

To add another level of complexity, let us consider the REM in a transverse field, i.e., the $p \rightarrow \infty$ limit of Eq.~\eqref{eq:quantum_p_spin_hamiltonian} in which the classical energy levels become independent and a finite-$\epsilon$ ``cluster'' corresponds to a single configuration.
Now $P_0$ is the projector onto configurations with classical energy density $\epsilon$, $V = - \Gamma \sum_i \hat{\sigma}_i^x$, and we calculate $\braket{\sigma | H_{\textrm{eff}} | \sigma'}$ for $\ket{\sigma}, \ket{\sigma'} \in P_0$.
Let the distance between $\sigma$ and $\sigma'$ be $x$.
$\braket{\sigma | H_{\textrm{eff}} | \sigma'} \sim O \big( \Gamma^{Nx} \big)$, and although there are many more terms at $Nx$\textsuperscript{th} order than in the toy problem, the same arguments hold here.
Each surviving term has an operator string of the form $V_{od} {V_d}^{Nx - 2} V_{od}$, with the intermediate states constituting a sequence of spin-flips transforming $\sigma$ into $\sigma'$ (a ``path'' in configuration space).
This gives us Eq.~\eqref{eq:forward_scattering_expression} in the main text.
The only subtlety is that some paths may pass through other states in $P_0$, giving operator strings $V_{od} {V_d}^{Ny - 2} {V_{od}}^2 {V_d}^{N(x-y) - 2} V_{od}$.
However, such paths are an exponentially small fraction of the total, and thus negligible: out of the $(Nx)!$ paths from $\sigma$ to $\sigma'$, the expected number with at least one intermediate configuration having energy density $\epsilon$ scales as $N e^{-N \epsilon^2} (Nx)!$~\footnote{By Markov's inequality, a typical sample cannot have more than a finite factor times the expected number of paths as $N \rightarrow \infty$.}.
Note that these atypical paths do not have amplitudes large enough to compensate for the smaller quantity, as the Schrieffer-Wolff formalism ensures that there are no resonant denominators.

\subsection{Time evolution}

Since the Schrieffer-Wolff transformation is unitary, time evolution from $\ket{\sigma}$ to $\ket{\sigma'}$ through $H$ is equivalent to time evolution from $e^S \ket{\sigma}$ to $e^S \ket{\sigma'}$ through $H_{\textrm{eff}}$.
Keep in mind that $H_{\textrm{eff}}$ is block-diagonal with respect to $P_0$ and $Q_0$.
Therefore $e^S \ket{\sigma}$ consists of two components, one that evolves within $P_0$ and one that evolves within $Q_0$.
Furthermore, $S$ is off-block-diagonal since every term in its expansion involves an odd number of $V_{od}$ factors (see Eq.~\eqref{eq:generator_expansion_orders}).
Thus $\cosh{(S)} \ket{\sigma}$ is the $P_0$ component and $\sinh{(S)} \ket{\sigma}$ is the $Q_0$ component.
In the main text, we simply take $e^S \ket{\sigma} \sim \ket{\sigma}$, and time-dependent perturbation theory in $V_{\textrm{eff}}$ then gives the rate at which the system tunnels between clusters.
However, to justify this, we must consider two effects which are not described by time-dependent perturbation theory.
First, $\cosh{(S)} \ket{\sigma}$ has amplitude not only in the initial cluster but in other clusters as well.
The overlap with final states vanishes at $t = 0$, since $\braket{\sigma' | e^{-S} e^S | \sigma} = 0$, yet this is due to interference between the terms in the expansion of $e^S$.
Such interference presumably decoheres by $t \sim O(1)$, meaning that the system may develop significant amplitude in other clusters on $O(1)$ timescales.
Second, $\sinh{(S)} \ket{\sigma}$ has non-zero weight, corresponding to probability for the system to be excited to higher classical energy densities.
The dynamics within the $Q_0$ subspace cannot be described by our method either.
In this subsection, we develop the conditions under which one can neglect these two effects.

Consider $\cosh{(S)} \ket{\sigma}$.
We want to calculate the amplitude on $\ket{\sigma'} \in P_0$.
We will continue to work within the FSA, meaning that we compute $\braket{\sigma' | \cosh{(S)} | \sigma}$ to lowest non-zero order in $\Gamma$.
By the same arguments as above, the relevant terms again have the operator string $V_{od} {V_d}^{Nx-2} V_{od}$, which all come from $\frac{1}{2} S^2$ in the expansion of $\cosh{(S)}$.
Unlike above, however, there are many more terms: the string in the first factor of $S$ must begin on $\ket{\sigma}$ but can terminate on any intermediate $\ket{\sigma''}$, and the string in the second factor must then begin at $\ket{\sigma''}$ and terminate on $\ket{\sigma'}$.
Thus
\begin{equation} \label{eq:other_cluster_amplitude_expression}
\braket{\sigma' | \cosh{(S)} | \sigma} \sim \frac{1}{2} \sum_{\sigma''} \sum_{P_{\sigma''}} \prod_{\sigma''' \in P_{\sigma''}} \frac{\Gamma}{N \big( \epsilon - \epsilon(\sigma''') \big) }.
\end{equation}
The outer sum is over all $\ket{\sigma''}$ intermediate between $\ket{\sigma}$ and $\ket{\sigma'}$.
The inner sum is over paths $P_{\sigma''}$ that pass through $\ket{\sigma''}$.

Focus on the REM for simplicity.
Then we again take $\epsilon(\sigma''') \rightarrow \mathbb{E} \big[ \epsilon(\sigma''') \big] = 0$, and
\begin{widetext}
\begin{equation} \label{eq:other_cluster_amplitude_evaluation}
\big| \braket{\sigma' | \cosh{(S)} | \sigma} \big| \sim \int_0^x N \textrm{d}y \, \binom{Nx}{Ny} \big( Ny \big) ! \big( N(x-y) \big) ! \left( \frac{\Gamma}{N |\epsilon|} \right) ^{Nx} = Nx \big( Nx \big) ! \left( \frac{\Gamma}{N |\epsilon|} \right) ^{Nx} \sim Nx e^{-N \gamma(x, \epsilon)},
\end{equation}
\end{widetext}
with $\gamma(x, \epsilon)$ as in the main text.
The extra factor of $Nx$ does not modify the exponential scaling and can be neglected.
To obtain the \textit{total} weight on other clusters, we multiply $e^{-2N \gamma(x, \epsilon)}$ by $e^{Ng(x, \epsilon)}$ and integrate over all $x \in \big[ x^{**}(\epsilon) , 1 - x^{**}(\epsilon) \big]$.
The result is $e^{-N r(\epsilon)}$ with
\begin{equation} \label{eq:total_other_cluster_amplitude}
r(\epsilon) = \min_{x \in [x^{**}(\epsilon), 1 - x^{**}(\epsilon)]} \big[ 2 \gamma(x, \epsilon) - g(x, \epsilon) \big] .
\end{equation}
Interestingly, the total weight is governed by the same exponent as the Fermi's golden rule rate, Eq.~\eqref{eq:tunneling_timescale}.
The field strength required for the transformed state $\cosh{(S)} \ket{\sigma}$ to have significant weight on other clusters, which could then decohere and become observable on $O(1)$ timescales, is exactly the field strength required for the tunneling rate to become $O(1)$ regardless.
If the tunneling rate is $O(1)$, our use of perturbation theory is questionable anyway.
Thus as long as we focus on the portion of the phase diagram for which $r(\epsilon) > 0$, it is justified to take $\cosh{(S)} \ket{\sigma} \sim \ket{\sigma}$.

Next consider $\sinh{(S)} \ket{\sigma}$.
We compute $\braket{\sigma' | \sinh{(S)} | \sigma}$ with $\ket{\sigma'} \not \in P_0$.
Thus $\epsilon(\sigma') \equiv \epsilon' \neq \epsilon$.
The lowest-order operator strings are of the form ${V_d}^{Nx-1} V_{od}$, which come from $S$ in the expansion of $\sinh{(S)}$. There is only one such term in Eq.~\eqref{eq:generator_expansion_orders}, giving
\begin{equation} \label{eq:excitation_amplitude_expression}
\braket{\sigma' | \sinh{(S)} | \sigma} \sim \sum_P \prod_{\sigma'' \in P} \frac{\Gamma}{N \big( \epsilon - \epsilon(\sigma'') \big) }.
\end{equation}
The notation is the same as for Eqs.~\eqref{eq:forward_scattering_expression} and~\eqref{eq:outside_amplitude_expression} in the main text.
For the REM, we evaluate Eq.~\eqref{eq:excitation_amplitude_expression} as before and obtain Eq.~\eqref{eq:outside_amplitude_evaluation}.
The conditions under which the total weight of $\sinh{(S)} \ket{\sigma}$ is negligible then follow as discussed in the main text.

\end{document}